\documentclass[12pt,preprint]{aastex}
\begin{document}
\def\hh{\, h^{-1}}
\newcommand{\wth}{$w(\theta)$}
\newcommand{\xir}{$\xi(r)$}
\newcommand{\Lya}{Ly$\alpha$}
\newcommand{\Lyb}{Lyman~$\beta$}
\newcommand{\Hb}{H$\beta$}
\newcommand{\HI}{H{\sc I}}
\newcommand{\Msun}{M$_{\odot}$}
\newcommand{\sfr}{M$_{\odot}$ yr$^{-1}$}
\newcommand{\sfrd}{M$_{\odot}$ yr$^{-1}$ Mpc$^{-3}$}
\newcommand{\cld}{erg s$^{-1}$ Hz$^{-1}$ Mpc$^{-3}$}
\newcommand{\dnsty}{$h^{-3}$Mpc$^3$}
\newcommand{\za}{$z_{\rm abs}$}
\newcommand{\ze}{$z_{\rm em}$}
\newcommand{\cmtwo}{cm$^{-2}$}
\newcommand{\nhi}{$N$(H$^0$)}
\newcommand{\degpoint}{\mbox{$^\circ\mskip-7.0mu \,$$~$}}
\newcommand{\halpha}{\mbox{H$\alpha$}}
\newcommand{\hbeta}{\mbox{H$\beta$}}
\newcommand{\hgamma}{\mbox{H$\gamma$}}
\newcommand{\kms}{\,km~s$^{-1}$}      
\newcommand{\minpoint}{\mbox{$'\mskip-4.7mu \mskip0.8mu \,$$~$}}
\newcommand{\mv}{\mbox{$m_{_V}$}}
\newcommand{\Mv}{\mbox{$M_{_V}$}}
\newcommand{\Mh}{M_h}
\newcommand{\peryr}{\mbox{$\>\rm yr^{-1}$}}
\newcommand{\secpoint}{\mbox{$''\mskip-7.6mu \\,$}}
\newcommand{\sqdeg}{\mbox{${\rm deg}^2$}}
\newcommand{\squig}{\sim\!\!}
\newcommand{\subsun}{\mbox{$_{\twelvesy\odot}$}}
\newcommand{\et}{{\it et al.}~}
\newcommand{\er}[2]{$_{-#1}^{+#2}$}
\def\h50{\, h_{50}^{-1}}
\def\hbl{km~s$^{-1}$~Mpc$^{-1}$}
\newcommand{\lseq}{\mbox{\raisebox{-0.7ex}{$\;\stackrel{<}{\sim}\;$}}}
\newcommand{\gseq}{\mbox{\raisebox{-0.7ex}{$\;\stackrel{>}{\sim}\;$}}}
\newcommand{\fesc}{f_{\rm esc}}
\newcommand{\fescrel}{f_{\rm esc,rel}}
\def\spose#1{\hbox to 0pt{#1\hss}}
\def\simlt{\mathrel{\spose{\lower 3pt\hbox{$\mathchar"218$}}
     \raise 2.0pt\hbox{$\mathchar"13C$}}}
\def\simgt{\mathrel{\spose{\lower 3pt\hbox{$\mathchar"218$}}
     \raise 2.0pt\hbox{$\mathchar"13E$}}}
\def\arcs{$''~$}
\def\arcm{$'~$}
\def\see{\mbox{$^{\prime\prime}$}}
\newcommand{\wu}{$U_{300}$}
\newcommand{\wb}{$B_{435}$}
\newcommand{\wv}{$V_{606}$}
\newcommand{\wi}{$i_{775}$}
\newcommand{\wz}{$z_{850}$}
\newcommand{\hmpc}{$h^{-1}$Mpc}
\newcommand{\um}{$\mu$m}

\title{The Great Observatories Origins Deep Survey:  Constraints on the
  Lyman Continuum Escape Fraction Distribution of Lyman--Break Galaxies
  at ${\mathbf{3.4<{\it z}<4.5}}$\altaffilmark{1}}


\author{\sc E. Vanzella\altaffilmark{2},
            M. Giavalisco\altaffilmark{3},
            A. Inoue\altaffilmark{4},
            M. Nonino\altaffilmark{2},
            F. Fontanot\altaffilmark{2},
            S. Cristiani\altaffilmark{2},
            A. Grazian\altaffilmark{5},
            M. Dickinson\altaffilmark{6},
            D. Stern\altaffilmark{7},
            P. Tozzi\altaffilmark{2},
            E. Giallongo\altaffilmark{5},
            H. Ferguson\altaffilmark{8},
            H. Spinrad\altaffilmark{9},
            K. Boutsia\altaffilmark{5},
            A. Fontana\altaffilmark{5},
            P. Rosati\altaffilmark{10}}

\affil{$^{2}$INAF -- Trieste Astronomical Observatory, via G.B. Tiepolo 11, 40131 Trieste, Italy}
\affil{$^{3}$Department of Astronomy, University of Massachusetts, Amherst MA 01003, USA} 
\affil{$^{4}$College of General Education, Osaka Sangyo University, 3-1-1, Nakagaito, Daito, Osaka 574-8530, Japan}
\affil{$^{5}$INAF - Rome Astronomical Observatory, Via Frascati 33, I-00040 Monteporzio Roma, Italy}
\affil{$^{6}$NOAO, PO Box 26732, Tucson, AZ 85726, USA}
\affil{$^{7}$Jet Propulsion Laboratory, California Institute of Technology, Mail Stop 169-527, Pasadena, CA 91109, USA}
\affil{$^{8}$STScI, 3700 San Martin Dr., Baltimore, MD 21218, USA}
\affil{$^{9}$Department of Astronomy, University of California, Berkeley, CA
  94720, USA}
\affil{$^{10}$ESO, Karl Schwarzschild Strasse 2, 85748, Garching, Germany}

\altaffiltext{1}{Based on observations made at the European Southern
  Observatory Very Large Telescope, Paranal, Chile (ESO programme 170.A-0788
  The Great Observatories Origins Deep Survey: ESO Public Observations of the
  SST Legacy / {\it HST} Treasury / {\it Chandra} Deep Field South). Also based on
  observations obtained with the NASA/ESA {\it Hubble Space Telescope}
  obtained at the Space Telescope Science Institute, which is operated by the
  Association of Universities for Research in Astronomy, Inc. (AURA) under
  NASA contract NAS 5-26555.}

\begin{abstract} 
We use ultra-deep ultraviolet VLT/VIMOS intermediate-band and VLT/FORS1
narrow-band imaging in the GOODS Southern field to derive limits on the
distribution of the escape fraction ($f_{\rm esc}$) of ionizing radiation for
$L \ge L^{*}_{z=3}$ Lyman Break Galaxies (LBGs) at redshift 3.4--4.5. Only one
LBG, at redshift $z=3.795$, is detected in its Lyman continuum (LyC;
S/N$\simeq$5.5), the highest redshift galaxy currently known with a direct
detection. Its ultraviolet morphology is quite compact ($R_{\rm eff}$=0.8\,
kpc physical).  Three out of seven AGN are also detected in their LyC,
including one at redshift $z=3.951$ and $z_{850} = 26.1$.  From stacked data
(LBGs) we set an upper limit to the average $f_{\rm esc}$ in the range
5\%--20\%, depending on the how the data are selected (e.g., by magnitude
and/or redshift).  We undertake extensive Monte Carlo simulations that take
into account intergalactic attenuation, stellar population synthesis models,
dust extinction and photometric noise in order to explore the moments of the
distribution of the escaping radiation.  Various distributions (exponential,
log-normal and Gaussian) are explored.  We find that the median $f_{\rm esc}$
is lower than $\simeq$6\% with an 84\% percentile limit not larger than
20\%. If this result remains valid for fainter LBGs down to current
observational limits, then the LBG population might be not sufficient to
account for the entire photoionization budget at the redshifts considered
here, with the exact details dependent upon the assumed ionizing background
and QSO contribution thereto.  It is possible that $f_{\rm esc}$ depends
 on the UV luminosity of the galaxies, with fainter galaxies having higher
 $f_{\rm esc}$, and estimates of $f_{\rm esc}$ from a sample of faint LBG
 from the HUDF (\wi$\le 28.5$) are in broad quantitative agreement with such
 a scenario.

\end{abstract}

\section{Introduction}
The fraction of the metagalactic ionizing background contributed
by star--forming galaxies remains poorly constrained  by direct
measures at every cosmic epoch because of the difficulty of the
observations. In particular, we do not have direct empirical
determinations of how the fraction of escaping ionizing radiation
depends on the properties of the galaxies, nor how it evolves with
redshift. Yet, the issue deserves attention, because it directly
bears on fundamental problems of galaxy evolution, such as the
evolution of the initial mass function (IMF) and the contribution
of galaxies to cosmic re-ionization.

The latter problem is currently particularly timely since observations
are starting to identify relatively large samples of galaxies at
$z>7$ (e.g., Bouwens et al. 2010a,b; Finkelstein et al. 2010;
Castellano et al. 2010), namely during the epoch when cosmic
re-ionization is believed to have completed (e.g., Fan et al. 2002).
The ultraviolet background (UVB) radiation can significantly affect galaxy
evolution by photo-ionizing and heating the interstellar medium
(ISM) to $\sim 10^{4}$\,K thereby decreasing gas accretion onto
low--mass galaxies and evaporating the existing gas in small
haloes. Deriving empirical constraints to the nature and evolution
of the cosmic UVB, as well as the nature of ionizing sources, remains
a primary goal of many observations.

Faucher-Gigu\`ere et al. (2008a) analyzed the opacity of the Lyman
alpha forest (LAF) of 86 high--resolution quasar (QSO) spectra and
found that the hydrogen photoionization rate $\Gamma$ is remarkably
flat in the redshift range 2 -- 4.2. The quasar contribution to the
hydrogen ionizing background increases toward $z\sim2$ as the peak
of the quasar luminosity function is approached (e.g., Hopkins et
al. 2007); beyond redshift 2 their contribution significantly
decreases (e.g., Fontanot et al. 2007; Siana et al. 2008;
Faucher-Gigu\`ere et al. 2009; Prochaska et al. (2009)).  
Glikman et al. (2010) calculate
the faint-end slope of the QSO luminosity function and find that
quasars might be able to ionize the intergalactic medium at $z
\simeq 4$. However, recent additional observations improve their
constraints on the slope, bringing it into greater agreement with
previous estimates and suggesting that QSOs {\em may not} be
sufficient to account for the ionizing photons (Glikman et al., in
preparation).  Star--forming galaxies are now known to exist
numerously at these redshifts and are therefore the leading candidates
to account for the remaining ionizing photons (e.g., Siana et al.
2008; Faucher-Gigu\`ere et al. 2009).

From the theoretical point of view,  current predictions of the
escape fraction of ionizing photons from high--redshift galaxies
are confusing, with different results obtained by different
simulations.  For example, Gnedin et al. (2008) argued that $f_{\rm
esc}$, i.e., ratio of the flux density of Lyman continuum (LyC)
escaping from a galaxy to that produced in the galaxy, increases
with increasing halo mass in the range of $\Mh = 10^{10} - 10^{12}$
\Msun, and their values of $f_{\rm esc}$ are mostly less than a few
per cent.  This is much lower than other published work; for
example, Wise \& Cen (2009) predict $f_{\rm esc} \sim 0.4$.  Yajima
et al. (2010) found an opposite behavior such that $f_{\rm esc}$
decreases with increasing halo mass, with an average $f_{\rm
esc}=0.40$ for $\Mh = 10^{9}$ \Msun\ dropping to $f_{\rm esc}=0.07$
for $\Mh = 10^{11}$ \Msun. A similar result was also found  by
Razoumov \& Sommer-Larsen (2010). It is clear that the physical
processes that modulate the escaping ionizing photons are not well
understood.

From the observational point of view, $f_\mathrm{esc}$ has been
poorly constrained due to the fact that LyC photons are easily
absorbed by both the IGM and the interstellar medium in a galaxy.
The best way to investigate the LyC emissivity from high--redshift
galaxies is to perform  deep spectroscopic or narrow-band observations
focused on the peak of the LyC emission, e.g., 880--910\AA\ rest-frame.
Ultra-deep intermediate-band imaging can also give an important
contribution, as we show in this work.

The LyC measure has been addressed in recent years by several
authors.  Malkan et al. (2003) and Siana et al. (2007, 2010) stacked
tens of deep ultraviolet images of galaxies at $z \sim$ 1
and report no detection.  Similarly, Cowie et al. (2009) combined
$\sim$ 600 galaxies at $z\sim1$ observed with {\it GALEX} and also
report a nondetection.  At higher redshift, Steidel et al. (2001)
initially found $f_{\rm esc,rel} \simgt 0.5$ from the composite
spectrum of 29 Lyman Break Galaxies (LBGs) at $z\sim 3$, where
$f_{\rm esc,rel}$ is the {\it relative} fraction of escaping LyC
(900\,\AA) photons relative to the fraction of escaping non-ionizing
ultraviolet (1500\,\AA) photons.  Giallongo et al. (2002) and Inoue et al.
(2005) estimated an upper limit of $f_{\rm esc,rel} \lseq 0.1-0.4$
for a sample of LBGs at $z\sim 3$.  Shapley et al. (2006; S06
hereafter) directly detected the escaping ionizing photons
from two LBGs in the SSA22 field at $z=3.1$, and estimated the
average value of $f_{\rm esc,rel}= 0.14$.  Chen et al. (2007) placed
a 95\% confidence level upper limit of 0.075 for the escaping radiation at
$z\ge2$ of star--forming regions hosting gamma-ray bursts.  More
recently, Iwata et al. (2009) detected the LyC emission from 10
Ly-$\alpha$ emitters (LAEs) and 7 LBGs within a sample of 198 LAEs
and LBGs in the SSA22 field.  They showed that the mean value of
$f_{\rm esc,rel}$ for the 7 LBGs is 0.11 after correcting for dust
extinction, and 0.20 if IGM absorption is taken into account.

Current observations suggest that $f_{\rm esc}$ increases with
increasing redshift: the fraction of direct LyC detection grows
from 0 to $\sim$10\% over the redshift range 0 $<z<$ 3 (e.g., Inoue
et al. 2006).  Even though the trend is possibly present, the current
fraction of direct LyC detections may be overestimated due to
contamination by blue light coming from lower redshift sources
superimposed on the targeted LBG. This has been investigated in
detail by Vanzella et al. (2010b), exploiting the high quality data
of the Great Observatories Origins Deep Surveys (GOODS; Giavalisco
et al. 2004a) and {\it Hubble} Ultra Deep Field projects (HUDF;
Beckwith et al. 2006) in conjunction with ultra-deep VLT/VIMOS
$U$--band imaging (Nonino et al. 2009).  They find that the probability
that at least $\sim$ 1/3 of the direct detections reported in the
literature are due to superposition of lower redshift sources
(confused in the PSF of the image) is larger than 50\%.  Therefore
the observed evolution of $f_{\rm esc}$ with redshift may be less
pronounced than currently believed.

It is therefore necessary to perform LyC measurements as free as
possible from contamination  by lower redshift sources. 
An ideal starting point
is therefore
deep, high resolution, multi-wavelength (space-based) imaging.
In the present
work we address this issue exploiting the extensive information
(spectroscopy and photometry) available in the GOODS Southern field
and the HUDF.  In particular, we take advantage of the deep VLT/FORS1
7$^{\prime}$$\times$7$^{\prime}$ narrow-band 3880\AA\ imaging
centered in the HUDF and ultra-deep intermediate-band VLT/VIMOS
$U$-band imaging of the entire GOODS-South (Nonino et al. 2009).

Throughout this paper magnitudes are reported in the AB scale (Oke 1974),
and the world model, when needed, is a flat universe with density
parameters $\Omega_m=0.3$, $\Omega_{\Lambda}=0.7$ and Hubble
constant $H_0=73$ \hbl.

\section{Data and sample selection}

\subsection{Intermediate-band (IB) imaging}
Ultra-deep $U$ intermediate-band imaging in the GOODS-South field was
performed with the VLT/VIMOS imaging spectrograph for a total
integration time of $\sim$ 40 hours.  Nonino et al. (2009) described
the reduction and characterization of the final image quality, which
reaches a depth of magnitude 29.5, 29.1 and 28.6 at 2$\sigma$,
3$\sigma$ and 5$\sigma$ within 1.2\see~aperture diameters,
respectively (see Table~\ref{tab:mags}).  Completeness and detection
limit analyses have been performed by running Monte Carlo simulations
and we refer the reader to Nonino et al. (2009) for details.  The
seeing of the co-added image is $\simeq$ 0.8\see~and represents the
deepest image currently available in the $U$ band. The depth and
the overall image quality of the co-added data, $\simeq$ 30 AB at
1$\sigma$, are well matched to the impressive multiwavelength data
available in GOODS-South.

\begin{figure}
 \epsscale{0.75}
 \plotone{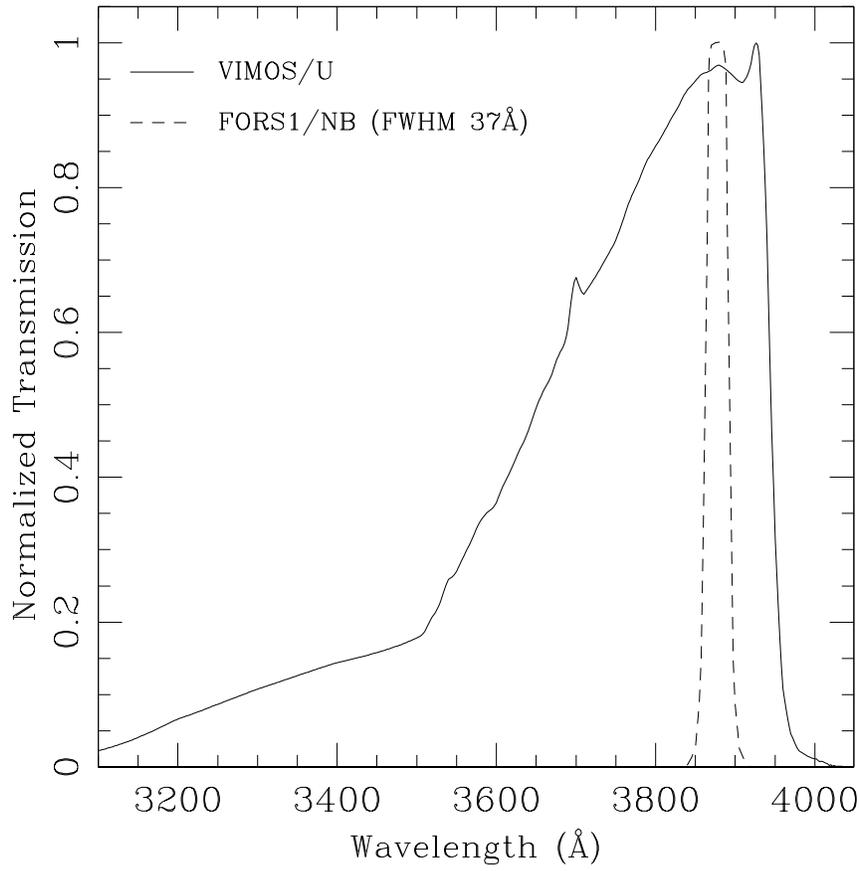} 
\caption{Normalized transmissions of the VIMOS/$U$ and narrow-band FORS1/3880 filters.\label{Fig1a}}
\end{figure}

The transmission of the filter is shown in Figure~\ref{Fig1a}.   The filter
probes the LyC region ($\lambda< 912$\AA) for sources at redshift
higher than 3.386.  In the following we only consider sources with
redshift higher than 3.4, for which the Lyman limit is beyond the
red limit of the filter. The transmission at $\lambda>$(912\AA$\times$4.4)
decreases rapidly to zero, and is never higher than 1\% of its peak
at $\simeq$ 3900\AA.  The filter has a FWHM of $\sim$ 350\AA,
corresponding to 80--60\AA~rest-frame for redshift 3.4--4.5, which
makes it an intermediate-band filter (IB, hereafter).  While the
lower limit of the redshift range investigated in this work is set
by the filter transmission, the upper limit is given by the gradual
increase of opacity of the IGM.  Indeed, as we discuss in detail
in Sect. 4.1, the average transmission of the IGM decreases as
redshift increases, reaching a transmission smaller than $3\times
10^{-4}$ (1.0 means 100\% transmission) at redshift beyond 4.5.
Therefore the transparency is too small at higher redshift to make
analysis of $z > 4.5$ galaxies useful. In the following we adopt
redshift 4.5 as an upper limit.

\subsection{Narrow-band (NB) imaging}
Very deep VLT/FORS1 narrow-band imaging (NB hereafter) has been
performed in the GOODS-South field, including the HUDF, centered at
$\alpha$=3$^{h}$32$^{m}$32$^{s}$.88,
$\delta$=$\mathrm{-}$27$^{d}$47$^{m}$16$^{s}$ (J2000)
 with a total exposure time of 60,900 seconds.  These data were
 obtained with the goal of detecting Lya emission at $z=2.2$
(see Hayes et al. 2010 for details).  The filter has a central
wavelength ($\lambda_{C}$) of 3880\ \AA, a width (FWHM) of $\Delta
\lambda$=37\ \AA, and is sensitive to the LyC region for galaxies
with redshift higher than 3.300. For the redshift range $3.3<z<4.5$,
the NB filter probes rest-frame wavelengths 902\ \AA~$>\lambda>$
700\ \AA.

The data were reduced using standard tasks in NOAO/IRAF, including
bias subtraction, flat field correction, and sky subtraction.  Images
were then registered onto a common astrometric grid and co-added.
The resulting magnitude limit of $\sim$ 26.5 at 5$\sigma$ within a
aperture diameter of 2\see~and the median seeing of the final image
of 0.85\see~are fully consistent with the reduction of Hayes et al.
(2010).  In particular, the NB image reaches the magnitude limit
of $\sim$ 29.0 at 1$\sigma$ within a 1.2\see~diameter aperture (see
Table~\ref{tab:mags}). 
The observed
field is a sub-region of the larger IB imaging, and therefore the
available LBG sample with spectroscopic redshifts is smaller ($\simeq$
1/4 of the full sample used in the VIMOS IB image).  However, useful
constraints can be derived from a stacking analysis (see Sect. 5).
In the following we mainly exploit the deeper and wider IB imaging.

\subsection{The spectroscopic sample}

Extensive spectroscopic redshift surveys have been performed in the
GOODS-South and surrounding fields (e.g., Cristiani et al. 2000; Szokoly
et al. 2004; Vanzella et al. 2006, 2008; Popesso et al. 2008;
Balestra et al. 2010; Stern et al., in preparation). A collection
of the published surveys is available at the ESO web site\footnote{{\tt
http://www.eso.org/sci/activities/projects/goods/}}.  In the present
work, only sources with secure redshifts are considered, i.e., those
with the highest quality. All the spectra and the identified spectral
features have been visually inspected.

The ESO/VIMOS spectroscopic survey extends beyond the deep GOODS-South
area, where the IB photometry is also available. In this extended
region we find 13 galaxies with secure redshifts in the range 3.4
$<z<$ 4.5.

\begin{figure}
 \epsscale{0.8}
 \plotone{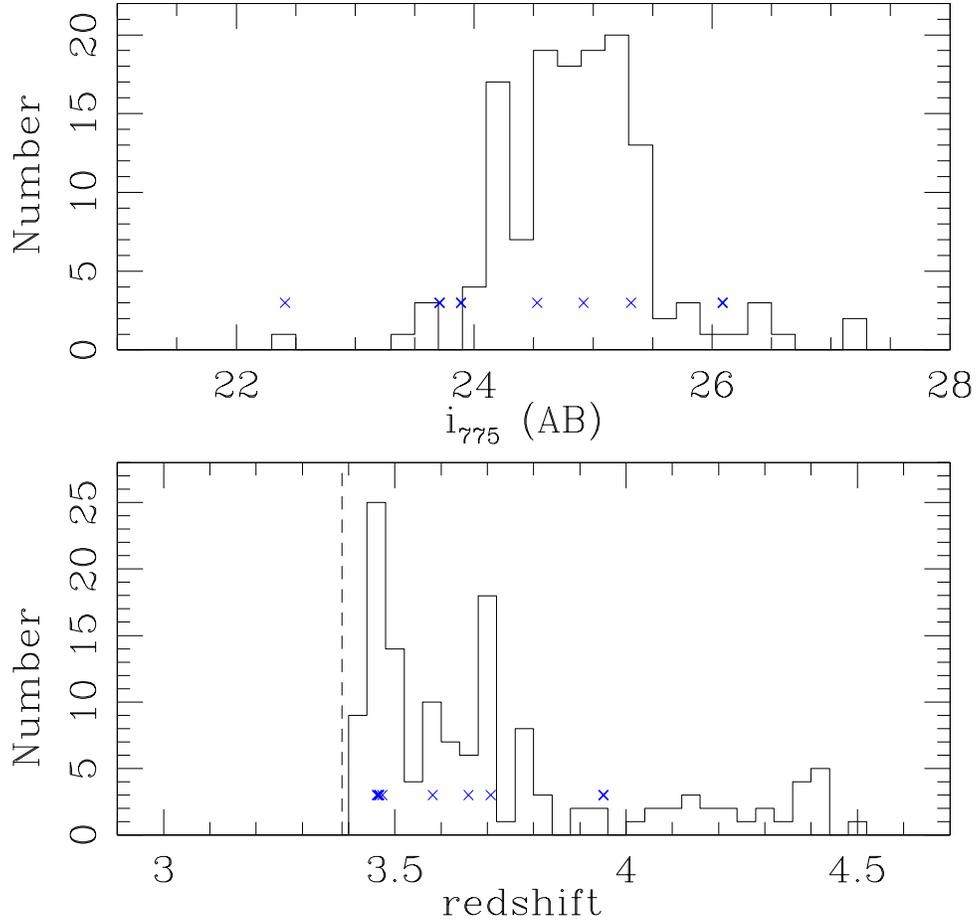}
\caption{Magnitude (top) and redshift (bottom) distributions of the 135 sources considered in this paper (7 AGNs and 128 galaxies).
The vertical dotted line in the bottom panel illustrates the minimum redshift probed by the IB
for the 912\AA~limit. In both panels blue crosses mark the AGNs. Those detected in their Lyman
continuum are indicated in bold face (see also Figure~\ref{AGNs}).   \label{MagZspec}}
\end{figure}

In total, 135 sources in the IB image (122 in the GOODS-South area and
13 outside) have secure redshifts in the range 3.4 $<z<$ 4.5. Their
redshift and \wi\ magnitude distributions are shown in
Figure~\ref{MagZspec}.  The mean redshift and \wi\ magnitude of the
sample are 3.64$\pm$0.27 and 24.85$\pm$0.58, respectively.

\section{The IB photometry and selection of the $clean$ sample}

Aperture photometry in the IB image was performed with SExtractor (Bertin \&
Arnouts 1996) in ``double image'' mode.  To detect sources we have
  created a new, fake image based on the IB one with pixel values set to zero
  anywhere except at the positions of the LBGs satisfying our selection
  criteria, which were set to 10,000.  Using these positions, photometry was
  then performed on the IB image. The accuracy in the centering of the
apertures has been tested on a sample of 68 spectroscopically identified stars
with magnitude \wz\ in the range $21-25$ uniformly distributed across the IB
image. The comparison between the original coordinates in the GOODS-South ACS
catalog (v2.0) and those obtained by SExtractor on the corresponding
``forced'' positions in the IB image shows a mean deviation of $\langle
\Delta_{\rm RA} \rangle$=0.001$\pm$0.133\see~ and $\langle \Delta_{\rm DEC}
\rangle$=$\mathrm{-}$0.009$\pm$0.136\see, which is significanly smaller than 1
pixel (0.3\see).  Flux measurements within increasing aperture diameters of
1.2\see,1.5\see,1.8\see~and 2.1\see~(4, 5, 6, 7 pixels) have been computed.
The same procedures have been executed for the NB image.

\begin{figure}
 \epsscale{1.0}
 \plotone{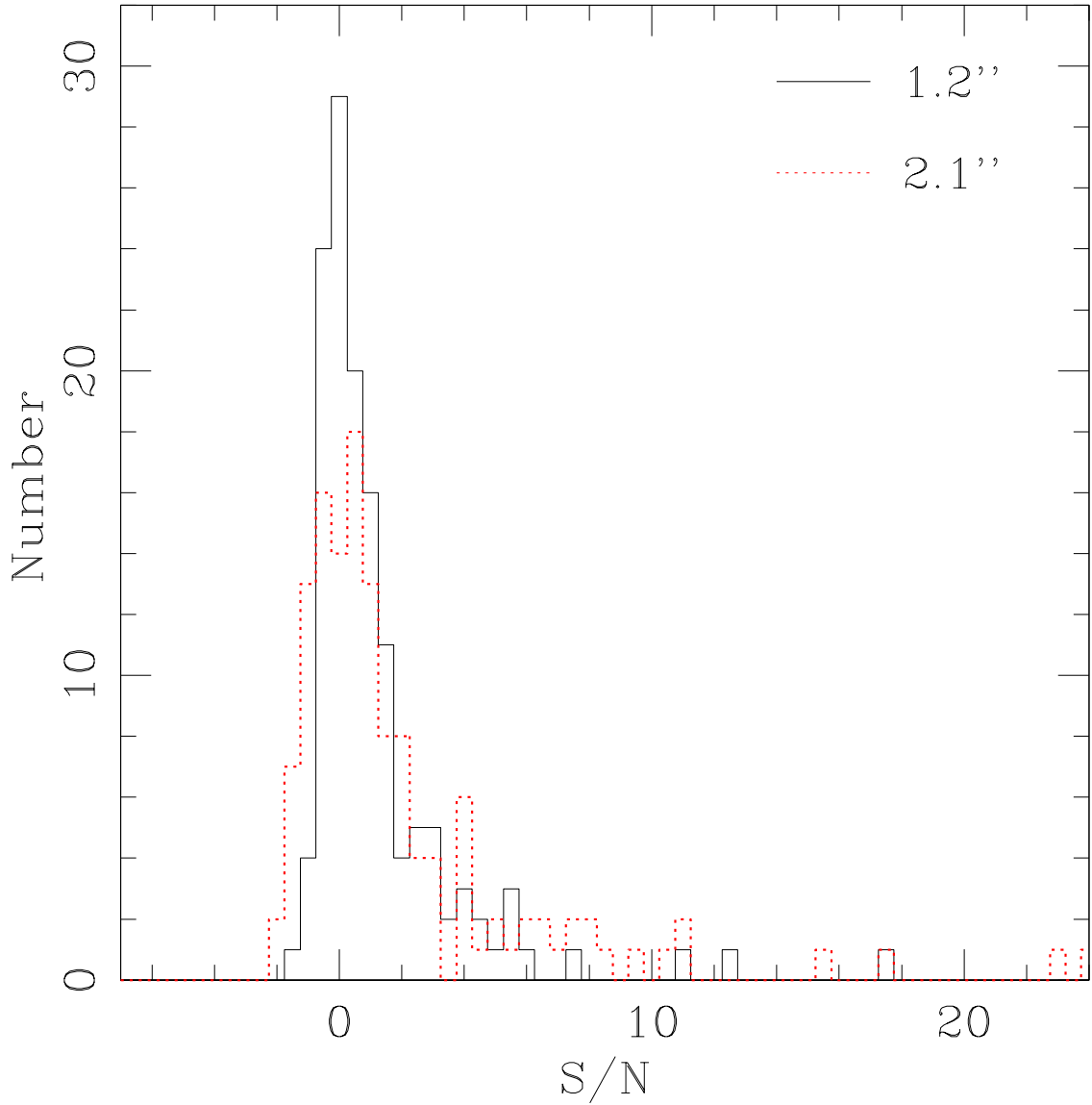} 
\caption{IB signal-to-noise ratio distributions calculated for the sample of 135 sources (7 AGNs and 128 galaxies).
Two out of four apertures are shown, 1.2\see~(solid-black line) and 2.1\see~(dotted-red line). The maximum of the distributions
peak around zero with a positive tail mainly due to intercepted foreground sources that fall partially in the aperture.
\label{Fig5}}
\end{figure}

Figure~\ref{Fig5} {shows the distribution of the IB signal-to-noise
(S/N) ratios for the 135 galaxies in 1.2\see~and 2.1\see~
apertures.} The distributions are asymmetric, peaked around zero,
and have a wider dispersion for the larger aperture. In the positive
tails of the distributions there are possible direct LyC detections
or intercepted foreground blue sources that mimic ionizing emission
(we refer to the latter as $contaminants$, see next section).  35
out of 135 sources have a S/N ratio in the IB image higher than 2
in either the 1.2\see~or 2.1\see~aperture.  All have been visually
inspected in the IB and the {\it HST} images (see Figures~\ref{ACS_IB1},
~\ref{ACS_IB2} and ~\ref{extended} and Appendix A for a description
of the sample in the outer region of the GOODS-South area).  The majority
are due to offset faint or bright sources that boost the flux measure
in the aperture centered on the LBG.  In these cases the S/N ratio
increases as the aperture diameter increases, because the contribution
of the nearby source also increases. Relatedly, if the signal arises
from the center of the aperture (i.e., at the LBG position), the
S/N typically decreases as the aperture size increases.
Illustrative examples of clear foreground contamination 
by bright, lower redshift galaxies include J033217.39-274142.4, J033212.98-274841.1, 
J033225.16-274852.6 and J033238.87-274908.7.  Examples with a 
distinct and offset faint, blue source clearly visible in the ACS images 
that significantly (if not totally) contribute to the aperture photometry include J033204.87-274451.4, 
J033220.97-275022.3, J033226.49-274124.0 and J033236.83-274558.0; the last one is in the HUDF and 
was discussed in Vanzella et al. (2010b; see Figure~\ref{Fig4}). 
Since we generally do not have the redshift of these faint, blue compact sources, it is not
possible to guarantee that they are in the foreground. However, we note that
the number of faint, nearby sources 
is consistent with the expected superposition probability (see next section).

In the following, the 1.2\see~apertures are used to derive
constraints on the escaping LyC radiation from LBGs. Moreover the
two sigma limit has been adopted as the main IB detection threshold
(results are also presented for 3 and 5 sigma limits).  We identify
27 out of 135 sources that most probably suffer contamination by
an offset foreground source in the 1.2\see~apertures. They are
excluded in the following analysis.  However, it is worth noting
that we would tend to underestimate the derived constraints on
$f_\mathrm{esc}$ if some of these offset sources are not foreground
contamination.


We are most interested in investigating the 
contribution of stellar emission to the UVB.  Therefore, AGNs are excluded from the sample 
as identified using either the 2 Ms {\it Chandra} image of GOODS-South (Luo et al. 2008) or by looking for 
typical AGN features like \ion{N}{5}, \ion{Si}{4} and \ion{C}{4}
emission lines in the spectra.
We find that 7 out of the 135 sources are AGNs, one of which is 
contaminated by a nearby, foreground source (e.g., is one of the 27 sources mentioned above).
The AGN image cutouts, photometric and spectroscopic information
are reported in Figure~\ref{AGNs}. Ingoring the contaminated
source, three out of the six remaining AGNs are detected in the IB image
with a S/N$>$2 (two with S/N$\sim$3 and one with S/N$\sim$2), i.e.,
at wavelengths bluer than 896\AA~rest-frame.  For
the highest redshift source, GDS~J033238.76-275121.6 at $z=3.951$,
the IB samples the rest-frame interval 700\AA-808\AA .

In summary, among the 135 sources (122 in the GOODS-South area and 13
outside), 128 are LBGs and 7 are AGNs.  Twenty-seven sources are contaminated
(26 LBGs and one AGN); the distribution of IB flux densities of the
uncontaminated sources is shown in Figure~\ref{Fluxes}. Of the 102
isolated LBGs, 92 are from the GOODS-South area and 10 are from the
sourrounding region (see Figure~\ref{CLEANpano}).  Images of the
26 contaminated LBGs (23 in the GOODS-South area and 3 in the outer
region) are shown in Figures~\ref{extended} and \ref{BADpano} (see
Table~\ref{tab:spec} for a summary).

The sample of 102 $clean$ LBGs is used to constrain the ionizing
radiation escape fraction.  In the following section we briefly
discuss the expected likelihood of foreground superposition that
can contaminate LyC measurements.

\subsection{Foreground contamination}
Vanzella et al. (2010b) discuss in detail the role of foreground
contamination in estimating the LyC radiation from galaxies at
redshift higher than 3.  Taking advantage of the ultra-deep imaging
available in the GOODS-South field, they show that the probability of
a foreground source mimicking LyC emission is not negligible. For
example, there is a 50\% chance that at least 15\% of a given sample
is affected by superposition by lower redshift sources for 1\see~seeing
and a $U$-band magnitude limit of 28.5 (Vanzella et al. 2010b).
Comparisons with the observations of Steidel et al. (2001) and
Shapley et al. (2006) have been performed using Monte Carlo
simulations.  Taking this contamination effect into account (which
increases with redshift), Vanzella et al. (2010b) estimates that
the escape fraction might be overestimated (amplified) by up to a
factor of two.

\begin{figure}
 \epsscale{0.85}
 \plotone{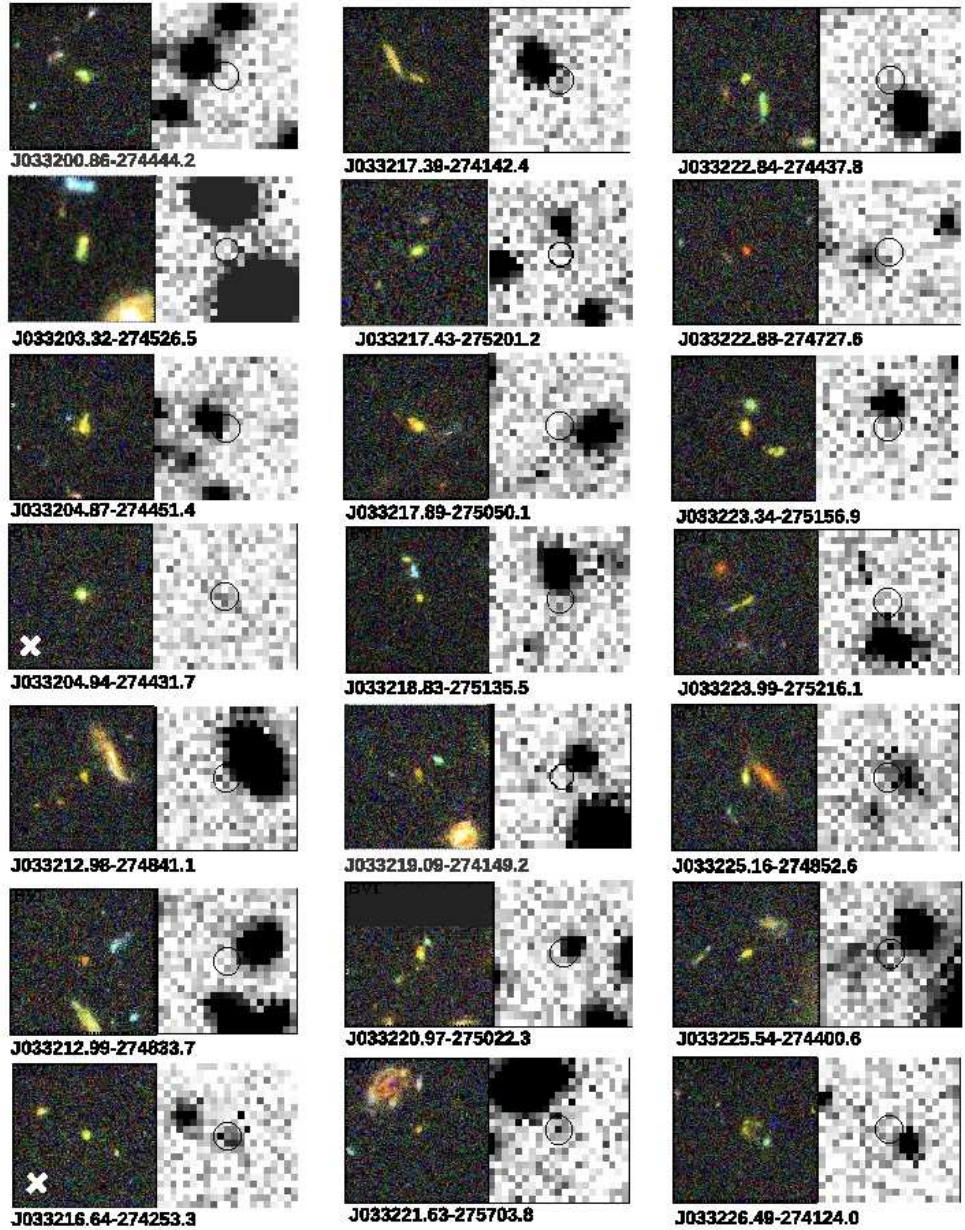} 
\caption{The {\it HST}/ACS $BVi$ color images and the ultra-deep VLT/VIMOS IB cutouts are shown for the sample with a detection above 
two sigma in the 1.2\see~or 2.1\see~apertures. The box sizes are 6.3\see~on a side and the circles show the 1.2\see~diameter aperture.   
Sources with a white cross are detected in the LyC region; the IB detections for the other
sources are all considered due to offset foreground contamination. \label{ACS_IB1}}
\end{figure}

\begin{figure}
 \epsscale{0.85}
 \plotone{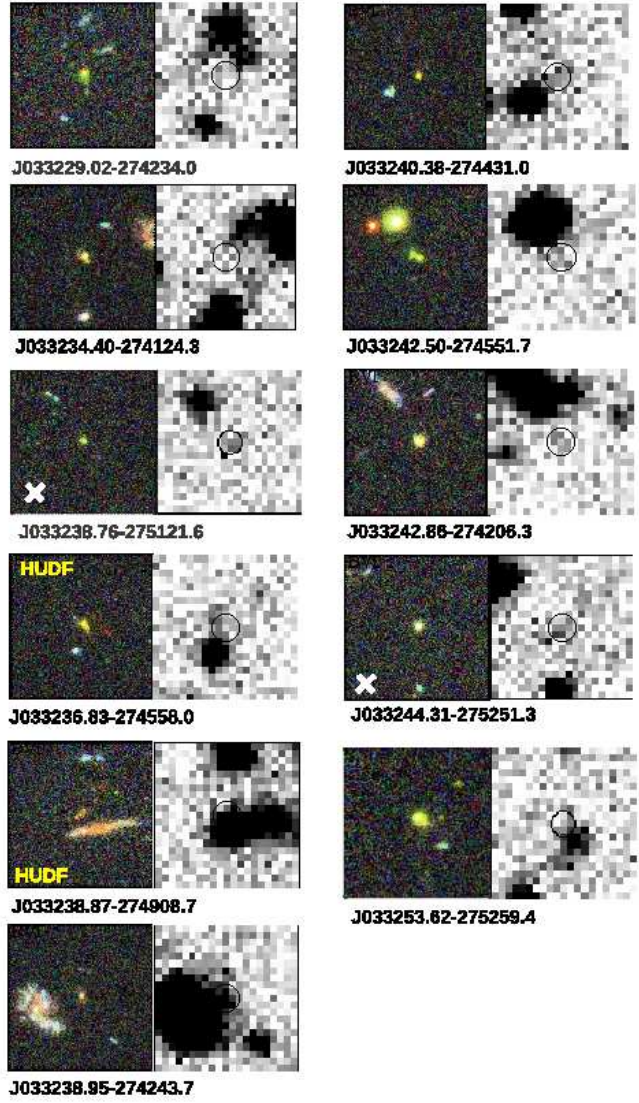} 
\caption{The same as Figure~\ref{ACS_IB1}. Objects belonging to the HUDF are indicated in the $BVi$ cutouts and
are shown at the HUDF depth in Figure~\ref{Fig4}. \label{ACS_IB2}}
\end{figure}

In this work we find contamination by both bright and faint sources
(Figures~\ref{ACS_IB1} and ~\ref{ACS_IB2}).  Considering the present
spectroscopic sample of 135 sources, including the 7 AGNs, we find
27 sources (one AGN and 26 LBGs) are contaminated by lower redshift
interlopers.

The probability that at least 13 high-redshift galaxies out of 135
are confused with a foreground object in a circle of 0.8\see~radius
and $U$-band magnitude down to 29.5 is $\sim$ 50\% (adopting the
$U$-band  number counts reported in Nonino et al. 2009 and Vanzella
et al. 2010b). However, the present analysis finds 27 contaminations
out of a sample of 135. The apparent inconsistency with the above
calculation is solved if the size of the nearby sources is taken
into account. Indeed, we clearly note from ACS and IB images 
that many extended
foreground galaxies still pollute the photometry of the background
LBG at separations even larger than 1\see. Looking carefully at
Figure~\ref{BADpano} (and Figures~\ref{ACS_IB1} and ~\ref{ACS_IB2})
it is apparent that $\simeq$ 12 out of 27 offset IB detections arise
from relatively close, compact blue sources at separations of $\sim$
1\see.  This is fully consistent with the expected probability of
a close superposition. The other superpositions are associated with
tails of extended galaxies at larger separations.

If we relax the above calculation and adopt a circle of radius
1.2\see~to calculate the interloper rate, the probability that at
least 27 galaxies out of 135 are polluted is $\sim$ 63\%. A dedicated
analysis should be performed to include the effect of size in these
calculations, but that is beyond the scope of the present work.
The main aim here is to select a sample as $clean$ as possible and
provide constraints on the escaping ionizing radiation.

\subsection{The LyC detections}
Four sources have been detected in their LyC (see
Table~\ref{tab:sampleIB}).  Three out of four are AGNs, two of them
are in the Szokoly et al. (2004; S04) spectroscopic catalog and
one has been observed in Cristiani et al. (2000; C00) and Balestra
et al. (2010).  The remaining source is an LBG observed with the
Keck-DEIMOS spectrograph (Stern et al., in preparation). Summarizing
the LyC detections,

\begin{enumerate}
\item{J033204.94-274431.7 : AGN. \ion{C}{4} and \ion{N}{5} emission lines are detected. No X-ray signal 
is measured in the 2Ms {\it Chandra} observations.}
\item{J033216.64-274253.3 : LBG. \Lya, \ion{Si}{4} and \ion{C}{4} (faint) absorption lines are detected.}
\item{J033238.76-275121.6 : AGN. \ion{C}{4}, \ion{C}{3}], \ion{C}{2} and X-ray emission are detected.}
\item{J033244.31-275251.3 : AGN. \Lya\ , \ion{N}{5}, \ion{Si}{4} and X-ray emission are detected.}
\end{enumerate}

\subsubsection{LyC emission from the LBG GDS~J033216.64-274253.3}
Among the 102 LBGs in the $clean$ sample, only one is detected in the IB
image (GDS~J033216.64-274253.3 at $z=3.795$; detected with S/N$\simeq$5.5). The GOODS ACS
images show that the source is quite compact, yet well resolved (SExtractor
stellarity index of 0.43 in the \wz\ band) with effective radius $R_{\rm eff}$=0.8 kpc 
physical ($R_{\rm eff}$=0.114\see), and has blue rest-frame ultraviolet continuum
(\wi$\mathrm{-}$\wz)=$\mathrm{-}$0.015 ($\beta$=$\mathrm{-}$2.1). There are no
close sources in the ACS images that might affect the IB signal.  Since it is
isolated and compact, the probability that another compact foreground source
is superposed along the line of sight within a circle of radius $R_{\rm eff}$ 
is lower than 0.1\%.  This is the highest redshift LBG currently known with
direct LyC detection.  The Keck-DEIMOS spectrum (Figure~\ref{LBGdetect}) shows
a clear \Lya\ break with a mean continuum decrement $D_{A}$=0.61$\pm$0.03,
consistent with the expected IGM transmission at redshift 3.8 (e.g., Inoue \&
Iwata 2008). The spectrum also shows faint \ion{Si}{4} and \ion{C}{4} absorption lines.
Interestingly, the low and high ionization absorption lines are weak (or
absent), in contrast to typical LBG spectra where weak interstellar absorption
lines are often associated with strong \Lya\ emission, and, conversely, \Lya\
in absorption is often accompanied with strong interstellar abosrption lines
(e.g., Shapley et al. 2003; Vanzella et al. 2009; Balestra et al. 2010). This
source appears to be the fortuitous combination of a relatively
high escape fraction of ionizing radiation with low IGM attenuation. From the
multi-wavelength information (MUSIC catalog, Grazian et al. 2006; Santini et al. 2009) 
we derive the following best-fit parameters for this galaxy using the libraries of
Bruzual \& Charlot (2003) and a Salpeter IMF (similar values are obtained using the libraries of
Charlot \& Bruzual 2007): extinction A1500 $\sim 0.62$ (assumuing
a Calzetti et al. (2000) extinction law), age $\simlt 0.1$ Gyr,
star formation rate SFR $\sim 26\ M_{\odot}{\rm yr}^{-1}$ and stellar mass 
$M_\star \sim 2.7\times10^{9}\ M_{\odot}$.
On the one hand, if we assume an IGM transmission of 100\%, a lower
limit of 15\% is obtained for $f_\mathrm{esc}$.  On the other hand,
an $f_\mathrm{esc}$ of 100\% corresponds to an IGM transmission not
lower than 0.15. We note that in this extreme case, no \Lya\ in
emission is expected as all of the ionizing radiation escapes. 
Moreover, the lower limit on the trasmission of 0.15 is higher than
the expected average value at this redshift, 0.022, and the probability 
to have a transmission higher than 0.15 at z=3.8 varies between 
4.5\% and 8\% (see simulations in Sect. 4). This indicates a line of sight 
particularly free from Lyman limit systems.

\subsubsection{LyC emission from AGNs and their influence on the IB photometry}
It is worth noting that the three AGNs with LyC detections are not
likely to be altering the transmission of the IGM in their spatial
proximity, including the volume probed by our $U$--band images.
Following D'Odorico et al. (2008; see also Cen \& Haiman 2000), we
calculated the radius of the sphere of influence (or Str\"omgren
sphere) of each of the detected AGN by relating the intensity of
the ultraviolet ionizing background at the Lyman limit to the luminosity of
the source at the same frequency. The resulting radius is smaller
than 750 kpc (physical) for all three AGNs. In particular, the faintest
and highest redshift of them, J033238.76-275121.6 at $z=3.951$, is
detected at S/N=3.3 in the IB image (i.e., $\sim$ 28.5) at a
rest-frame wavelength blueward of 808\AA. Its influence on the
surrounding IGM reaches a radius of only $\simeq$ 250 kpc (physical).
Indeed, the flux in the $U$--band would be completely suppressed if
a Lyman limit system was intercepted in the redshift range $3.386 - 3.951$
(i.e., in the wavelength interval between the red edge of the $U$
filter and the Lyman limit of the source). Since the Str\"omgren
sphere radius is only 250 kpc (physical), the AGN is not influencing
the IB observation, or in other words, the source must ionize the
IGM at least down to redshift 3.386 to perturb the IB photometry
($\Delta z \ge 0.56$), which clearly is not the case. Its LyC
detection is therefore most probably due to an intrinsically high
transmission of the IGM and/or escape fraction of ionizing radiation.

Similarly, if $f_\mathrm{esc}$ is intrinsically high for the LBGs
considered here as well, we expect a certain number of detections
in the ultra-deep IB image (see below). In other words, the LyC
detection of some of the AGNs validates the statistical method
adopted here in constraining the $f_\mathrm{esc}$ distribution for
galaxies.

The next section describes Monte Carlo simulations performed with
the aim of constraining the $f_\mathrm{esc}$ distribution. A deeper
limit on its average is given in Sect. 5 by stacking the sources.

\section{Constraining the distribution of escaping ionizing radiation}
In the following analysis we refer to the $clean$ spectroscopic
sample described in the previous sections, composed by 102 galaxies
with one LyC detection at S/N$\simeq$5.5.  Once
this $clean$ spectroscopic sample of LBGs has been identified, it
is interesting to address the following question: how many sources
do we expect to detect at a given depth in the IB  survey assuming
a distribution function of $f_{\rm esc}$?

Allowing redshift to vary from 3.3 to 4.5, the IB filter probes
rest-frame wavelengths far below 912\AA~(e.g., down to $\sim 700$\AA),
where the IGM transmission decreases rapidly to zero because of the
increasing probability of intercepting Lyman limit systems and
damped \Lya\ systems (hereafter LLSs and DLAs, respectively) as
well as the decreasing free path of ionizing photons (see next
section).  It is therefore necessary to estimate the expected LyC
signal in our IB image adopting a model of the IGM transmission.
This is also useful for the source stacking (Sect. 5).

\subsection{Modeling the IGM transmission}
The effective optical depth through a clumpy IGM at the rest-frame
frequency $\nu_{S}$ for a source at redshift $z_{S}$ is (e.g.,
Paresce, McKee \& Bowyer 1980):

\begin{equation}
 \tau_{\rm eff}(\nu_{\rm S}, z_{\rm S}) = \int_0^{z_{\rm S}} dz
  \int_{N_{\rm l}}^{N_{\rm u}} d N_{\rm HI}
  \frac{\partial^2 {\cal N}}{\partial z \partial N_{\rm HI}}
  (1-e^{-\tau_{\rm cl}})\,,
\end{equation}

\noindent where $\partial^2 {\cal N}/\partial z \partial N_{\rm HI}$ is the number
of absorbers along the line of sight per unit redshift $z$
interval and per unit HI column density $N_{\rm HI}$ interval, and
$\tau_{\rm cl}=\sigma_{\rm HI}(\nu_{\rm S}(1+z)/(1+z_{\rm S}))N_{\rm HI}$
is the optical depth of an absorber with $N_{\rm HI}$ at $z$,
where $\sigma_{\rm HI}(\nu)$ is the HI cross section at frequency
$\nu$ in the absorber's rest-frame.\footnote{The frequency dependence of $\sigma_{\rm HI}(\nu)$
for LyC is approximately $\propto \nu^{-3}$.}
If the column density distribution of the absorbers is a
power-law with index $-\beta$ ($\beta\approx1.5$; e.g., Kim et al. 2002)
independent of redshift, the maximum contribution to $\tau_{\rm eff}$
is made by absorbers with $\tau_{\rm cl}\sim1$. Therefore, the absorption of the Lyman
continuum is mainly caused by LLSs and DLAs
with $N_{\rm HI}>10^{17}$ cm$^{-2}$ and not by the LAF, which has
$N_{\rm HI}\sim10^{13}$ cm$^{-2}$. This implies that LyC absorption is very stochastic 
because it is related to the probability of intercepting a LLS.

In this work the intergalactic absorption derived from the Monte Carlo simulations of 
Inoue \& Iwata (2008) is adopted (IW08 hereafter). Briefly,  we recall the main steps.
The simulations are based on an empirical distribution function of intergalactic absorbers
which reproduces the observational statistics of the LAF, LLSs and DLAs
simultaneously. From this assumed distribution function, a large number of absorbers
have been generated (running suitable Monte Carlo simulations) along many lines of sight.
The probability to encounter an absorber is assumed
to follow a Poisson distribution, and for each one the column density and Doppler parameter
are extracted randomly from their (empirical) probability distribution functions.
Typically $\sim$ 18,000 absorbers are generated for a line of sight in the redshift 
interval $0<z<6$ (this number depends on the lower limit to the column density).
As described in detail in IW08, 10,000 lines of sight have been calculated in the redshift interval 
$3.4 \leq z \leq 4.5$ with step $\Delta z$=0.1. The resulting
mean intergalactic transmission is comparable to that derived by Meiksin (2006) in the Lyman
series regime ($\lambda>$ 912\AA), though the IW08 transmission are 
slightly lower in the LyC regime. This is due to the 
different number of LLSs considered by the two approaches (see IW08 for details).

Figure~\ref{TransSingle} shows examples of transmissions along different line of
sights, extracted randomly from the 10,000 realizations at the three redshifts
$z=3.4$, $3.7$ and $4.0$. In some cases the transmission drops to zero blueward of the redshift of LLSs;
in others the signal coming from the source is transmitted down 
to $\simeq$700\AA . In general, as redshift increases, the IB filter used here is strongly penalized.
This is shown in Figure~\ref{Trans}, where the medians and 68\% confidence interval of the transmissions of 10,000
different lines of sight calculated in the redshift range 3.4--4.5 are reported (the averages are also
shown as open squares).
The distributions are not symmetric because of intervening LLSs and DLAs. The transmissions
have been convolved with the IB filter shape; therefore, they are calculated in the suitable
wavelength interval covered by the filter at a given redshift. For comparison, the median transmissions
calculated in the wavelength range 880-910\AA~is also shown (it is identical to that reported in
Figure 8 of IW08). Clearly the transmissions
calculated through the IB filter are systematically lower than the ``optimal'' case (880-910\AA). 
This is fully taken into account in the simulations we describe in the following section.
We note that at redshifts beyond 4.0, the intergalactic absorption strongly attenuates the  
ionizing flux. 

We briefly note the recent findings of Prochaska et al. (2010), in which they find
a significantly lower incidence of LLSs at $z<4$ compared with previous estimates. 
A similar result have been found by Songaila \& Cowie (2010), even though this tendency is
less pronounced.
Qualitatively, if these results are correct then the transmission
of the IGM derived here is underestimated; i.e., the number of
expected LyC detections in our IB survey would increase, and given
the observational constraint of only one out of 102 LBGs detected,
this would imply that the upper limits we derive for $f_\mathrm{esc}$
are further strengthened (see next section).  Indeed, the fact that
we detect two sources in their LyC (one LBG and one AGN) at relatively
high redshift ($z\sim4$) may support a higher average transmission
than predicted from our simulations.

Quantitatively, the detailed inclusion of the results of Prochaska
et al. (2010) (and Prochaska et al. 2009; Songaila \& Cowie 2010)
in the modeling of the IGM (as in IW08) deserves a dedicated work
that will be presented elsewhere (Inoue et al. 2010).  
However, a comparison between
the observations of P\'eroux et al. (2005; e.g., those adopted in IW08),
Prochaska et al. (2010) and Songaila \& Cowie (2010), shows that
decreasing the mean LyC optical depth (due to a lower number density
of LLSs) increases the final IGM transmission by a factor of 1.5
(see Figure~\ref{new_lls}).  

In Sect. 5 we report limits on $f_\mathrm{esc}$
by stacking and considering previous and current statistics on
LLSs.

\subsection{Simulating the expected number of LyC detections}
The {\it relative} fraction of escaping LyC  photons (at 900\,\AA)
relative to the fraction of escaping non-ionizing ultraviolet (1500\,\AA)
photons is defined as (Steidel et al. 2001): \begin{equation} f_{\rm
esc,rel} \equiv \frac{(L1500/L900)_{\rm int}}{(F1500/F900)_{\rm
obs}} \exp(\tau^{\rm IGM}_{900}), \label{f_esc_rel} \label{eq:fesc_rel}
\end{equation}

\noindent where $(F1500/F900)_{\rm obs}$, $(L1500/L900)_{\rm int}$ and $\tau^{\rm IGM}_{900}$
represent the observed 1500\,\AA/900\,\AA\ flux density ratio,
the intrinsic 1500\,\AA/900\,\AA\ luminosity density ratio, and
the line-of-sight opacity of the IGM for 900\,\AA\ photons, respectively.
Equation~(\ref{eq:fesc_rel}) compares the observed flux density ratio
(corrected for the IGM opacity) with models of the ultraviolet spectral energy
distribution of star-forming galaxies.
If the dust attenuation $A_{1500}$ is known, 
$f_\mathrm{esc,rel}$ can be converted to $f_\mathrm{esc}$ as 
$f_\mathrm{esc} = 10^{-0.4A_{1500}} f_\mathrm{esc,rel}$
(e.g., Inoue et al. 2005; Siana et al. 2007).
We can rearrange the above equation to give an estimation of the observed flux 
at wavelengths smaller than the Lyman limit (i.e., $F_{\rm LyC}$ instead of $F900_{\rm obs}$):

\begin{equation}
F_{\rm LyC} =  \left ( \frac{L\lambda_{\rm rest}}{L1500} \right )_{\rm int}  f_{\rm esc} \times (F1500)_{\rm obs}\times e^{-\tau^{\rm IGM}_{\lambda}} \times 10^{0.4\times A1500}, \label{f900_obs}
\label{eq:f900_obs}
\end{equation}

\noindent The quantities on the right side of the equation have
been modeled and inserted in a Monte Carlo simulation. They are
described as follows:

\begin{enumerate}
\item{{$\bf \left (\frac{L1500}{L\lambda_{rest}}\right )_{int}$:} Depending on the redshift, 
the wavelength range probed by the IB filter is included in
the interval $\lambda_{\rm rest}<908$\AA~(=$\lambda_{\rm rest}({\rm max})$=4000\AA/(1+$z_{\rm min}$) 
with $z_{\rm min}$=3.405).
The value of the intrinsic luminosity density ratio $(L1500/L\lambda_{\rm rest})_{\rm int}$ is 
still very uncertain
observationally; it must therefore be estimated from stellar population synthesis models.
The LyC flux is emitted by O stars, whose lifetime is much shorter than the 
B and A stars that dominate the 1500\AA~flux emission. 
Therefore the luminosity ratio depends on the stellar population age, metallicity, star formation 
history (single burst, exponential decay, constant or multi-bursts) and IMF 
(e.g., Bruzual \& Charlot 2003; Leitherer et al. 1999).
When the dying O stars are not replenished with new star formation, the ratio increases
rapidly within a few million years (e.g., single burst). In the case of constant star formation 
rate (SFR), O stars are continuously formed
and the A and B stars accumulate, so the ratio slowly increases and saturates at later times, beyond 1 Gyr 
(Siana et al. 2007). Inoue et al. (2005), adopting the Starburst 99
models (Leitherer et al. 1999) and assuming a constant SFR, Salpeter IMF
over the mass range 0.1-100\Msun~and a metallicity Z of 0.001-0.02 (0.02 is the Solar value), obtained 
ratios that lie in the 1.5$<$$(L1500/L900)_{\rm int}$$<$5.5 interval. 
Depending on the time since the onset of star formation, they reported $(L1500/L700)_{\rm int}$=4.0 (7.0) 
in the case of 10\ Myr (100\ Myr) old stellar populations, with the  value saturating at older ages.
Here, wavelengths below 900\AA~are observed (down to $\simeq$750\AA) with the IB filter passing from redshift
3.4 to 4.5 and therefore a suitable ratio must be considered. Adopting an average age of $\simeq$300 Myr for
our sample (derived by Pentericci et al. 2007, 2010), and following the calculations of
Siana et al. (2007) that reported ratios between 6 and 8 for $(L1500/L900)_{\rm int}$ and 
$(L1500/L700)_{\rm int}$ for a
similar age, respectively, and Inoue et al. (2005) that reported a ratio $\simeq$ 7, 
we adopt a value of $(L1500/L_{\rm LyC})_{\rm int}$$\simeq$7 for the following analysis. 
This has been used in the Monte Carlo simulations described below, where we assume  
a Gaussian distribution with a mean of 7 and a standard deviation 50\% of the mean.
The 50\% scatter includes the dispersion due to different physical 
properties of the LBGs in the sample as well as their redshift distribution. Results do not change 
significantly if we allow it to vary between 30\% and 70\%.}

\item{{$\bf (F1500)_{obs}$:} The $(F1500)_{\rm obs}$ is derived from the observed \wi\ magnitude of each source. 
That filter corresponds 
to $\lambda_{\rm eff}$ $\sim$ 1750\AA, 1550\AA~and 1400\AA~at redshift 3.4, 4.0 and 4.5, respectively. 
The average spectral slope of the sample is almost flat, $\langle \beta \rangle=-1.95 \pm 0.4$,
so the estimated flux density deviates by only a small amount from the 
observed \wi\ magnitude (less than 5\% on average, $F_{\nu}$ $\sim$ $\lambda^{2-\beta}$).}

\item{{\bf A1500:} The correction for dust attenuation has been calculated assuming the empirical extinction
relation $A1500$ = 4.43 + $\beta$1.99, where the spectral index $\beta$ is derived from the 
observed (\wi $-$\wz) color following the prescription of Bouwens et al. (2009b;
see also Meurer et al. 1999). This technique has already been employed by several previous studies estimating 
the SFR density at $z$ $\sim$ 2--6 (e.g., Adelberger \& Steidel 2000; Meurer et al. 1999;
Bouwens et al. 2006; Stark et al. 2007). The dust correction has also been compared to the 
values derived from the standard SED fitting of a sub-sample of the 102 LBGs (80\% of them) for which we have
photometric multi-wavelength coverage from the MUSIC catalog. We refer the reader to Santini et al. (2009) for details 
of the SED fitting procedures. The median and standard deviation
of A1500 from the SED fitting is $0.61_{-0.61}^{+0.93}$, while from the ultraviolet spectral slope it is
$0.58_{-0.58}^{+0.88}$ assuming a Calzetti extinction law.}

\item{{\bf IGM attenuation:} The transmission of the IGM (T = $e^{-\tau^{\rm IGM}_{\lambda}}$) between the observer and 
the source redshift has been inserted by adopting the models $\tau^{\rm IGM}_{\lambda}$ of IW08
(see previous section).}

\end{enumerate}

A Monte Carlo simulation that takes into account all of the above quantities has been performed and is described 
next. Random IGM transmissions have been associated with each object from the spectroscopic sample (102 galaxies) 
by extracting from the 10,000 different lines of sight at the closest redshift to the source 
and convolved with the IB filter. Similarly, a value of the intrinsic ratio
of the luminosity density has been extracted randomly from the adopted distribution described above,
then a correction for the dust absorption (A1500) has been calculated from the observed color.
The $f_\mathrm{esc}$ has been investigated by inserting various functional behaviors (see next section).
An estimate of the flux $(F900)_{\rm obs}$ is derived from Eq.~\ref{f900_obs}. 
If the estimated $(F900)_{\rm obs}$ flux is brighter than the adopted threshold (i.e., the depth of the IB image), 
then it has been further perturbed according to the
error of the image photometry for that flux level. The error as a function of the magnitude has been parameterized 
analytically by fitting an exponential function to the observed data.

Ten thousand simulated samples of 102 galaxies, each anchored to the observed quantities of the $clean$ sample, i.e., the rest-frame ultraviolet colors, magnitudes and redshift,
have been generated for each $f_\mathrm{esc}$ distribution.
For each of the 10,000 extractions, the number of sources brighter than the chosen IB magnitude limit 
is recorded. At the end, for each assumed $f_\mathrm{esc}$ distribution, 10,000 estimations of the expected number 
of ``survived'' sources is calculated, and the median and central 68 percentile range are derived.

Summarizing, the number of expected LyC detections in the IB band has been calculated performing Monte Carlo simulations that take
into account the IGM transmission, the IB filter shape, the distribution of the intrinsic luminosity ratio ($L1500/LF_{\rm LyC}$),  
dust attenuation by the interstellar medium, photometric errors of the IB image, the observed redshift and \wi\ magnitude.

Once these effects are suitably modeled, the expected number of
LyC detections in the IB image depends on the moments of the $f_\mathrm{esc}$ distribution assumed.
The aim of the next section is to investigate these dependencies through comparison
with the observed number.

\subsection{The tested distributions}
It is reasonable to believe that $f_\mathrm{esc}$ varies from galaxy
to galaxy with a distribution currently not known.  Indeed, theoretical
studies propose various behaviors for $f_\mathrm{esc}$ as a function
of the halo mass, luminosity, gas and dust content, geometry, etc.
(Gnedin et al. 2008; Wise \& Cen 2009; Yajima et al. 2010; Razoumov
\& Sommer-Larsen 2010).  We have investigated which effect an assumed
distribution of  $f_\mathrm{esc}$ would have on the expected number
of LyC detections in our IB image.  We assume that the
distributions apply for all luminosities.

Before introducing the various functional forms adopted, we
perform a similar check to what was done by Siana et al. (2007) in
characterizing their null detection of LyC at $z\sim1.3$. We assume
a fraction (Y) of our sample has constant $f_\mathrm{esc}$ (=X) and
the rest (1-Y) has zero LyC emission (X and Y vary between 0 and
1).  Monte Carlo simulations have been run in order to estimate the
number of expected detections (N) in our IB survey as described in
the previous section down to a 2-sigma limit and as a function of
X and Y.  This has been done 30 times (1000 extractions each) on
the clean sample randomly sorted at every time. The results are
reported in Figure~\ref{check} where points (X,Y) with N=3 belong
to the black region; above it N$>$3 and below N$<$3.  In our sample
of 102 LBGs only one has been detected, N=1.  Very low $f_\mathrm{esc}$
($<$5\%) are needed to reproduce the null or one LyC detection if
all LBGs have the same $f_\mathrm{esc}$ value.  Conversely, a high
$f_\mathrm{esc}$ ($>$70\%) can reproduce a null or one detection
if it is associated with less than 10\% of the LBG sample.  This
test suggests that high values of $f_\mathrm{esc}$ are less probable
in this luminosity regime (a feature already noted in other works;
e.g., Giallongo et al. 2002; Inoue et al. 2005; Shapley et al. 2006;
Iwata et al. 2009).

Having this result in mind, various continuous functions have
been explored: flat, Gaussian and asymmetric functions (exponential
and log-normal).  In detail, we test the following functional forms
for $f_\mathrm{esc}$:

\begin{enumerate}
\item{{\bf Constant value:} A constant value of $f_\mathrm{esc}$
has been assumed for all galaxies, between 0.0 and 1.0 with an
increment of 0.01. It is not realistic to assume a constant value
of $f_\mathrm{esc}$. However it is useful as a check of the typical
scatter due to solely Monte Carlo simulated effects like IGM, dust,
photometric noise, intrinsic luminosity ratio distribution, etc.}

\item{{\bf Gaussian distribution:} $f_\mathrm{esc}$ is assumed to
be distributed as a Gaussian form with a mean running from 0.0 to
1.0 (step 0.01) and standard deviation equal to half of the mean.}

\item{{\bf Exponential distribution: } Exponential distributions
($e^{-\lambda}$) with different slopes $\lambda$ have been considered,
with $\lambda$ running from 1 to 100 with step $\Delta \lambda$=1.
This allows us to investigate the effect of asymmetric tails toward
high $f_\mathrm{esc}$ values.}

\item{{\bf Log-normal distribution:} Log-normal distributions with
various medians and scatter have been inserted, $e^{-K+\lambda
\times \rm Gauss}$, where $\rm Gauss$ is extracted randomly from a
Gaussian distribution with zero mean and standard deviation equal
to 1.  $K$ has been assumed to be 1, 2, 3, 4 and for each $\lambda$
parameter running from 0.1 to 10.0 (step 0.1). Varying $K$ allows
us to change the average of the initial symmetric distribution
(small $\lambda$). As $\lambda$ increases, the median of the
distribution tends to zero and an asymmetric tail toward high values
arises (see below).} 
\end{enumerate}

In this way 100 constant, Gaussian, exponential and 400 log-normal
distributions of $f_\mathrm{esc}$ have been calculated, each one
extracted 10,000 times with the Monte Carlo simulation described
above.  We discuss the results In the following section.

\subsection{Constraints on the ionizing radiation fraction distribution from the spectroscopic sample}
The expected number of LyC detections has been explored as a function
of the median and 68\%  interval of the assumed $f_\mathrm{esc}$
distributions.  Figure~\ref{MC} and \ref{MC_exp} show the results
for constant, Gaussian and exponential $f_\mathrm{esc}$ behaviors.
As expected, in all cases, if the median $f_\mathrm{esc}$ increases
the number of expected LyC detections also increases. Considering
the IB depth at the 2$\sigma$ level (left panels of Figure~\ref{MC})
and given the single LyC detection, the upper limit on the median
$f_\mathrm{esc}$ is $\simeq$ 6(5)\% at 3$\sigma$ for the Gaussian
(exponential) distribution. Relaxing to a shallower IB depth, 29.1
(3$\sigma$) and 28.6 (5$\sigma$), the median of the Gaussian
(exponential) $f_\mathrm{esc}$ distribution is lower than 12(10)\%
and 20(15)\%, respectively.

Focusing on the exponential distribution, 
a median less than $\sim$ 5\% and a scatter less than $\sim$ 15\% are required 
to be compatible within 3$\sigma$ to the observations. In other words, the very low number of
LyC detections in the IB image down to 29.5 limit (2$\sigma$ within 1.2\see~diameter) implies
an upper limit to the median $f_\mathrm{esc}$ and the 84\% percentile of the distribution
of 5\% and 15\%, respectively. Figure~\ref{MC_exp} shows in more detail the exponential case 
reported in the lower left panel of Figure~\ref{MC}, i.e., IB depth at 2$\sigma$ level. In particular,
the distributions that lead, on average, to an expected number of LyC detections higher than 4 are
highlighted in the [median -- scatter] plane (see inner right box of the same figure). 
Following the Poisson statistics, these are
distributions for which the probability to have less than two LyC detections (our case) is lower than 5\%.

While in the above cases the dispersion decreases together with the median of the distribution and approaches 
zero as the median tends to zero,
the effect of a relatively high scatter and very small median of $f_\mathrm{esc}$ has been explored 
by adopting log-normal distributions as described in the previous section.
The case with $K=1$ and $\lambda$ running from 0.1 to 10.0 ($\Delta \lambda = 0.1$) is shown in Figure~\ref{MC_LOGN}.
As $\lambda$ increases, the distribution changes its shape from symmetric and centered at the initial value $e^{-1}$ 
($K=1$) to asymmetric with very small median and relatively high 84\% scatter (see examples in the main box 
of Figure~\ref{MC_LOGN}).
In the extreme case of very small medians (e.g., $\lambda=100$, median $e^{-100}$) the
scatter still allows a marginal  LyC detection. 
This is the reason why in Figure~\ref{MC_LOGN} the expected number of LyC detections is different from zero 
even though the median is close to zero.
Similarly, for the exponential case, the locus of points in the [median -- scatter] plane 
of all the log-normal distributions (varying $\lambda$ and $K=1$) is shown in the 
inner right box of Figure~\ref{MC_LOGN}. Those excluded with a probability higher than 95\% 
have been highlighted. In this case, 
log-normal distributions with a scatter lower than $\sim$ 18\% are favoured
if compared with observations. Similar results have been found varying 
log-normal distributions with $K=1, 2, 3, 4$ and $\lambda$ running from 0.1 to 10.0
(see Figure~\ref{MC_LOGN_MED_SIG}). 

Summarizing, among the $f_\mathrm{esc}$ distributions explored and from the comparison with 
the observed number of LyC detections (i.e., 1 out of 102) we find that the median fraction of ionizing 
radiation escaping from the LBG sample considered here is less than $\sim$ 5-6\% with a 1$\sigma$ scatter
(upper eighty-fourth percentile) not larger than $\sim$20\% at the 2$\sigma$ IB depth.
These upper limits increase to $\sim$ 10-12\% and 20\% (median and  1$\sigma$) if the 3$\sigma$ IB 
depth is considered. In general, adopting the Poissonian statistics and considering the single LyC
here reported, the distributions that predict more than 5 (10) LyC detections can be excluded with a 
probability higher than 95\%(99\%), respectively.

\section{Upper limits on the ionizing radiation fraction from stacking}
\subsection{IB imaging}
The median and average stacking of all 102 galaxies are shown in the top part to Figure~\ref{STACK}. No LyC 
detection is seen. Similarly, the median and average stacks have been performed for the sub-sample of 45 LBG
with redshift lower than 3.6, for which the mean IGM transmission is higher (see Figure~\ref{Trans}).
Median and averages have also been calculated for two sub-samples of these 45 galaxies,
one of 22 LBG with \wi\ $<$ 25 and the other for 23 LBG with \wi\ $>$ 25 (they are shown
in Figure~\ref{STACK}). None of them show a LyC detection. It is worth noting that the individual 
LyC measure described in Sect. 3.2.1. (when included) does not provide enough counts to contribute to a 
significant stacked detection.

Following previous work in the literature (e.g., Steidel et al. 2001;
Giallongo et al. 2002; Inoue et al. 2005; Shapley et al. 2006; Iwata et al. 2009), 
an upper limit on the $f_\mathrm{esc}$ can be calculated 
by assuming average values for the quantities in Eq.~\ref{eq:fesc_rel} and correcting $f_\mathrm{esc,rel}$  
for the average dust extinction at 1500\AA. 
Figure~\ref{Fluxes} shows the IB flux (AB) distribution within the 1.2\see~aperture of the
sample of 102 LBGs plus 6 AGNs. Excluding the AGNs and the single LBG detected in its LyC, the distribution 
has a mean and standard deviation of $\langle$F(1.2\see)$\rangle$ = $\mathrm{-}$0.007$\pm$0.023 $\times$ 
$10^{-30}$ erg~sec$^{-1}$~cm$^{-2}$~Hz$^{-1}$. This one sigma dispersion corresponds to magnitude 30.50 AB,
consistently with the one sigma limit described at the beginning, and can be adopted as the typical error 
of the single measure (assuming the flux distribution to be Gaussian and each measurement
as independent). Therefore the one sigma limit of the mean over $N$ sources is 
0.023 $\times$ $10^{-30}$ erg~sec$^{-1}$~cm$^{-2}$~Hz$^{-1}$ decreased by the square root of $N$.
In the case of 102 LBGs, this limit corresponds to magnitude 33. However, despite the very deep flux reached, 
the very low transmission of the IGM as redshift increase ($z > 4$) weakens the constraints 
on $f_\mathrm{esc}$.
In order to keep a relatively high IGM transmission and high magnitude contrast,
galaxies with redshift lower than 3.75 and \wi\ $<$ 25.5 have been selected (64 LBGs).
The average \wi\ magnitude and redshift are 24.84 and 3.57, respectively, and the observed one sigma flux density
ratio probed is $(F1500/F\lambda_{rest})_{\rm obs}$ = 1473. The upper limit on $f_\mathrm{esc,rel}$  is
\begin{equation}
f_\mathrm{esc,rel} < \left [ \frac{7}{1473} \right ] \times\frac{1}{0.09} = 0.05,
\label{eq:stack1}
\end{equation}

\noindent where the average transmission is $\simeq$ 0.09 at redshift below 3.75 (see Figure~\ref{Trans}) 
and the intrinsic luminosity ratio has been fixed to 7 (see previous section).
Assuming an average A1500=0.65 (flat spectral slope), the upper limit on total escape fraction 
$f_\mathrm{esc} = f_\mathrm{esc,rel} \times 10^{-0.4A1500}$ turns out to be 0.03.



Several further constraints on $f_\mathrm{esc}$ can be calculated by selecting subsamples in magnitude 
and redshift. Selecting brighter sources allows one to increase the magnitude contrast,
and selecting lower redshift sources allows one to increase the average IGM transmission because the IB 
filter approaches rest-frame 900\AA . 
A summary of this is shown in Table~\ref{tab:fesc}
where upper limits on $f_\mathrm{esc}$ are reported as a function of the magnitude threshold (columns) and 
redshift (rows). 
The upper limits on the (mean) escaping ionizing radiation $f_\mathrm{esc}$ span the range 4\% to 60\%.
The values derived from the Monte Carlo simulations are consistent with this 
interval, in particular if the brighter (large magnitude contrast) and lower redshift (higher IGM transmission) 
objects are considered in the grouping (see Table~\ref{tab:fesc}).
Compared to the estimations appearing in the literature
so far, these are the most constraining results on $f_\mathrm{esc}$ in the redshift and magnitude range here 
considered.

As discussed at the end of Sect. 4.1, it is worth noting the
effect of the recent findings of Prochaska et al. (2010) and Songaila
\& Cowie (2010) about the lower incidence of LLSs observed with
respect to previous work (e.g., P\`eroux et al. 2005; also adopted
in this work). Assuming these results to be correct, they imply a
higher IGM transmission by a factor of 1.5 (this has
been done by rescaling the predictions of IW08 to match the number
of LLSs reported in those works).  Therefore, the limits we report
in the above equation would further decrease by the same factor
(Eq.~\ref{eq:stack1}).  The same dimming factor would also apply
to the upper limits reported in Table~\ref{tab:fesc}.


\subsection{NB imaging}

The number of available sources of the $clean$ sample in the FORS1 NB imaging is 30.
There are two more LBGs in this sample (not considered in the IB calculations) since the 
NB imaging starts to probe the LyC at redshift beyond 3.3 (not 3.4 as for the IB).
None of the LBGs show a LyC detection at S/N$>$2, and again no detection is measured 
in the median (average) stacking. 
While the NB observations have the disadvantage of having smaller statistics
and being shallower than the IB imaging, the narrow wavelength window helps 
to increase the average transmission of the IGM if a suitable selection in redshift
is done (the limits calculated from the whole sample do not add any further constraint 
with respect to the IB derivations). 
Indeed, selecting sources with redshift below 3.65, the average transmission 
turns out to be 0.2 and the probed rest-frame wavelengths span the interval
843-890\AA. As above,  an intrinsic luminosity ratio of 
$(L1500/L\lambda_{\rm rest})_{\rm int}$ = 7 is adopted (see Sect. 4.2).
Eight galaxies match the selection $3.3<z<3.65$ and \wi\ $<25.5$, with 
averages $\langle z \rangle$=3.5 $\pm$ 0.1 and $\langle$\wi$\rangle$=24.6 $\pm$ 0.4.
In the average process the one sigma lower limit of the flux ratio probed (magnitude NB$\sim$29) 
is $F1500/F_{\rm NB}$=161, which translates to an upper limit on $f_\mathrm{esc}$ of 12\% assuming 
an intrinsic ratio of 7 and the same dust extinction A1500 adopted for the IB stacking
(it is $f_\mathrm{esc,rel}$ of 22\%).
As discussed in the previous section adopting the correction factors for the
transmission of Songaila \& Cowie (2010) and Prochaska et al. (2010) these limits would be lowered.

\section{Evidence for a luminosity dependency?}

The limits on $f_\mathrm{esc}$ derived from Monte Carlo simulations and from the stacking have been calculated
from the spectroscopic 
sample of LBG, which is probing mainly $L \gseq L^{*}_{z=3}$ luminosities,
i.e., galaxies hosted by relatively massive dark matter halos ($\simeq$ $10^{11}M_{\odot}$; e.g., 
Arnouts et al. 2002, Lee et al. 2009).
If we adopt the $z \sim 4$ luminosity function of Bouwens
et al. (2007) integrated down to the faint limit $M_\mathrm{UV}=-16$
($0.02 L^{*}_{z=3}$), the one-sigma limit  $f_\mathrm{esc,rel}$$\simeq$5\% derived from the brighter
part of the sample here analyzed and assumed valid for all luminosities, 
we find the ionizing background intensity provided by the LBG population at $z\sim4$ to be
$J_{\nu}({\rm LyC})<0.8 \times 10^{-22}~{\rm erg}~{\rm s}^{-1}~{\rm Hz}^{-1}~{\rm cm}^{-2}~{\rm sr}^{-1}$.
\footnote{The ionizing luminosity density 
$\rho_{LyC}=\rho(1500)\frac{L_{\rm LyC}}{L1500}(f_\mathrm{esc,rel}$) is independent from the
intrinsic luminosity ratio, see Eq.~\ref{eq:fesc_rel}.}  This value represents $\simeq$ 16\% and
19\%  of the total ultraviolet background intensity measured by  Giallongo
et al. (1996) and Bolton et al. (2005), respectively, while it
increases to 40\% of the estimation of Faucher-Gigu\`ere et al.
(2008b). \footnote{Bolton et al. (2005) and Faucher-Gigu\`ere
et al. (2008) calculate the total photoionization rate $\Gamma_{-12}$
of 1.0 and 0.5, respectively (where $\Gamma_{-12}=\Gamma/10^{-12}{\rm s}^{-1}$).
We convert these values into intensity units 
(${\rm erg}~{\rm s}^{-1}~{\rm Hz}^{-1}~{\rm cm}^{-2}~{\rm sr}^{-1}$) by adopting the
formulation in Schirber \& Bullok (2003): $\Gamma_{-12}=(12/(3+\alpha_{\rm
UV}))\times J_{-21}$ and assuming an $\alpha_{\rm UV}$ spectral
slope of the background intensity equal to 2 (e.g., Haehnelt et al.
2001) and $J_{-21}$ is the intensity of the ultraviolet background in units
of $10^{-21}~{\rm erg}~{\rm s}^{-1}~{\rm Hz}^{-1}~{\rm cm}^{-2}~{\rm
sr}^{-1}$ (see Eq. 7 of Schirber \& Bullok 2003).} If we
integrate the luminosity function down to $M_\mathrm{UV}=-10$, these
fractions increase slightly to 23\%, 28\% and 56\%, respectively.
Therefore star-forming galaxies alone may not be able 
to account for the entire ultraviolet ionizing budget.  It is worth noting that,
given the observed global non-detection here reported, a higher
transmission of the IGM (Prochaska et al. 2010) would imply an even
tighter constraint on the escaping radiation and therefore less
contribution by (bright) LBGs. If galaxies only partially contribute,
the remaining fraction presumably is provided by QSOs. However,
several works agree on the fact that at redshift beyond 3, galaxies
play a dominant role in the IGM ionization with QSOs contributing
fractions of only 30--10\% in the redshift range 3.5--4.0, (e.g.,
Bolton et al. 2005; Siana et al.  2008; Faucher-Gigu\`ere et al.
2008b; Faucher-Gigu\`ere et al. 2009; Prochaska et al. 2009).
If galaxies are responsible for the remaining ionizing budget, then the
implication is that $f_\mathrm{esc}$ depends on the UV luminosity, with
fainter galaxies having a larger escape fraction. 

Such a relationship between $f_\mathrm{esc}$ and the UVluminosity has recently
been suggeted by simulations.  Yajima et al. (2010) performed a
three-dimensional radiation transfer calculation of stellar radiation for a
large number of high-redshift, star-forming galaxies in cosmological
simulations.  One of their primary conclusions was that, in the redshift
interval $3<z<6$, galaxies become the main contributor to IGM ionization with
the average (standard deviation) of the escape fraction of ionizing radiation
increasing to $\simeq$ 40\% (20\%) for low mass haloes, $\Mh<10^{10}$\, \Msun.
For the larger haloes, $\Mh\simeq10^{11}$\, \Msun, they predict an average
$f_\mathrm{esc}$ of 7\% with a relatively small scatter, less than 20\% (see
their Figure 2), in agreement with what we find here.

Sources at the fainter end of the magnitude distribution are
under--represented in our simulations, since the magnitude contrast reached
between the ionizing and non-ionizing radiation (IB and \wi\ ) is smaller.
However, if there is a dependency of $f_\mathrm{esc}$ with the luminosity so
that fainter galaxies have higher average $f_\mathrm{esc}$, this could
partially compensate for the lower contrast.  As mentioned above, Yajima et
al. (2010) predicts $f_\mathrm{esc}$ increases for lower halo masses, with
$\langle f_\mathrm{esc}\rangle$ of 40$\pm$30\% and 15$\pm$20\% for
$\Mh=10^{9}$ \Msun\ and $\Mh=10^{10}$ \Msun, respectively.  Razoumov \&
Sommer-Larsen (2010) predict $f_\mathrm{esc}$ values that reach 70-80\% for
$\Mh$ in the range $10^{9}$--$10^{10}$ \Msun\ at redshift 4.4.

To investigate the IB emission in our data at fainter limits we have selected
a sample of 218 galaxies with photometric redshifts in the range $3.4<z<4.0$
and magnitude $27<$ \wi $<28.5$ extracted from the public photometric redshift
catalog of Coe et al. (2006).  IB photometry has been performed at the
positions of the galaxies in the four aperture diameters, as was done for the
spectroscopic sample. Twenty-six out of 218 have a detection with S/N ratio
higher than 2 in the 1.2\see~diameter aperture; Figure~\ref{CLEANpanoBPZ}
shows their IB cutouts and the list is reported in
Table~\ref{sample_BPZ}. Five out of 26 show an IB emission aligned with the
LBG position (marked with black crosses in Figure~\ref{CLEANpanoBPZ}).  In the
other cases an offset is present and the LBG may suffer contamination by
foreground sources.

From Monte Carlo simulations of the sample considered here we find that the
expected median number of LyC detections at the 2$\sigma$ IB depth is
$6_{-2}^{+3}$ and $3_{-1}^{+2}$ for the case of constant and Gaussian
distributions with median $f_\mathrm{esc}$=1.0 and 0.7,
respectively. Exponential and log-normal distributions predicts a comparable
number ($\simeq3-6$) if the median of the $f_\mathrm{esc}$ distributions is
larger than 60\%.  This result is quantitatively consistent with the
expectations if $f_\mathrm{esc}$ increases for less luminous galaxies.

However, we stress that apart from the reduced magnitude contrast
probed, the main disadvantage concerning these fainter sources is
the reliance on photometric redshifts.  If the sample includes some
galaxies with true redshifts $z<3.4$, the IB image would include
light from wavelengths longward of the Lyman limit.

\section{Conclusions}

Exploiting the ultra-deep VIMOS IB and deep FORS1 NB imaging of the
GOODS-South field, new limits on the escape fraction of ionizing radiation
from star-forming galaxies at redshift 3.4--4.5 have been derived.
Particular care has been devoted to clean the spectroscopic
sample from foreground contamination and AGN contributions. From a
sample of 102 LBGs we derive the following results:

\begin{enumerate}
\item{From Monte Carlo simulations and stacking of the IB and NB
imaging, we find that $f_\mathrm{esc}$ of $L \ge L^{*}_{z=3}$ LBG
is distributed with a median lower than 5-6\% and 84 percentile
scatter lower than 20\% in all the distributions investigated
(Gaussian, exponential and log-normal). We note that the low upper
limit on the median escape fraction is for the entire sample,
independent of spectral properties. If the recent findings of
Prochaska et al. (2010) and Songaila \& Cowie (2010) are considered
--- i.e., the average IGM transmission is higher than that adopted
here --- then the limits we derive are further strengthened.}

\item{One star-forming galaxy is detected in its LyC region at
700--835\AA~rest-frame. It is the highest redshift galaxy with such
a detection currently known and is most probably due to a combination
of high IGM transmission coupled with a relatively high $f_\mathrm{esc}$.
The lower limit on $f_\mathrm{esc}$ is 15\%; assuming
$f_\mathrm{esc}$=100\%, the IGM transmission cannot be lower than
15\%. This value is higher than the expected average value at this
redshift (2.2\%), implying that it is a particularly free line of
sight. The galaxy shows a blue UV--continuum spectral slope
($\beta$=$\mathrm{-}$2.1) and weak or absent interstellar absorption
lines in the spectrum even though \Lya\ is in absorption. The
ultraviolet morphology is quite compact, $R_{\rm eff}$=0.8 kpc
(physical).} 
\end{enumerate}

Adopting the observed photoionization rate of Bolton et al. (2005) or
Faucher-Gigu\`ere et al. (2008b), star-forming galaxies contribute partially
($\simlt 50\%$) to the required ultraviolet ionizing budget if
$f_\mathrm{esc}$ is constant and equal to 5\%.  On one hand, the contribution
of QSOs may still be significant at the redshifts considered here, providing
the ionizing fraction missed by galaxies.  This strongly depends on the
faint-end slope of the QSO luminosity function (Glikman et al. 2010, and in
preparation).  On the other hand, several works suggest that the QSO
contribution to the UVB decreases significantly beyond redshift 2, reaching
fractions lower than 50\% (down to 10\%) at redshift 4 (e.g., Fontanot et
al. 2007; Siana et al. 2008; Faucher-Gigu\`ere et al. 2008b; Prochaska et al.
2009).  In this case galaxies would provide almost all of the ionizing
radiation, which, as we have seen, requires that $f_\mathrm{esc}$ depends on
the UV luminosity. We remind, however, that these conclusions depend on both
the total ionizing UVB and the QSO fractional contribution to it, quantities
remain empirically poorly constrained at these redshifts.

If $f_\mathrm{esc}$ does indeed depend on the UV luminosity, then we can
speculate on the following scenario. Bouwens et al. (2009b), analyzing samples
of LBGs in the redshift range $3<z<6$ show that there is a clear correlation
between the UV--continuum slope $\beta$ and ultraviolet luminosity.  In
particular, for the $B$--band dropout sample also adopted here, more luminous
LBGs have redder colors. Moreover, it is known from stacking tens and hundreds
of LBG spectra at redshift 3--5 that the redder UV-continuum slopes are linked
to low \Lya~equivalent widths and stronger interstellar absorption lines,
while Ly$\alpha$ emitters are bluer and have weaker interstellar absorption
lines (e.g., Shapley et al. 2003; Pentericci et al. 2007; Vanzella et
al. 2009; Balestra et al. 2010).  

Thus, if $f_\mathrm{esc}$ is, on average, larger in galaxies with fainter 
UV luminosity then we would expect that the bulk of the ionizing radiation
comes from faint \Lya\ emitters, which are, in general, younger and
less massive than their brighter LBG counterparts (e.g., Ono et al. 2010).  On
the one hand, this would be plausible and possible cases have been found by
Iwata et al. (2009).  On the other hand, we have shown an opposite example, in
which LyC emission arises from a LBG without \Lya\ in emission (even though it
has been detected in the bright $L^{*}_{z=3}$ regime).

A direct investigation at fainter flux limits (\wi$>$27) is challenging
because the magnitude contrast decreases and spectroscopic redshifts are
difficult to obtain with current facilities. An analysis of faint galaxies
from the HUDF, 27$<$\wi$<$28.5 or 0.3$L^{*}_{z=3}$ -- 0.04$L^{*}_{z=3}$, 
selected with photometric redshift is in broad quantitative agreement with the
expectations if $f_\mathrm{esc}$ depends on the UV luminosity, increasing for
fainter galaxies. 

A way to explore this faint luminosity regime (before the advent of future
telescopes like {\it JWST} and the ELTs) is to analyze samples of
spectroscopically confirmed LAEs selected through NB imaging (e.g., Iwata et
al. 2009; Inoue te al. 2010), looking at peculiar spectroscopic features 
related to low-luminosity AGN or hot and massive stars (e.g. Vanzella et al. (2010a))
or using spectra of $\gamma$-ray burst
afterglows (e.g., Chen et al. 2007), strategies that we plan to pursue in
upcoming works.

\acknowledgments
We would like to thank the anonymous referee, whose comments improved the paper.
We are grateful to the ESO staff in Paranal and Garching who greatly
helped in the development of this programme. 
We acknowledge financial contribution from contract ASI/COFIN I/016/07/0 and PRIN INAF 2007
``A Deep VLT and LBT view of the Early Universe''. 
We would like to thank D. Schaerer for useful discussions about the single
LBG detection and E. Glikman for the recent estimates of the QSO luminosity 
function at the faint end at redshift 4.
The work of DS was carried out at Jet Propulsion Laboratory, California 
Institute of Technology, under a contract with NASA.

\appendix
\section{The sources in the extended GOODS-South region}
\begin{figure}
 \epsscale{0.8}
 \plotone{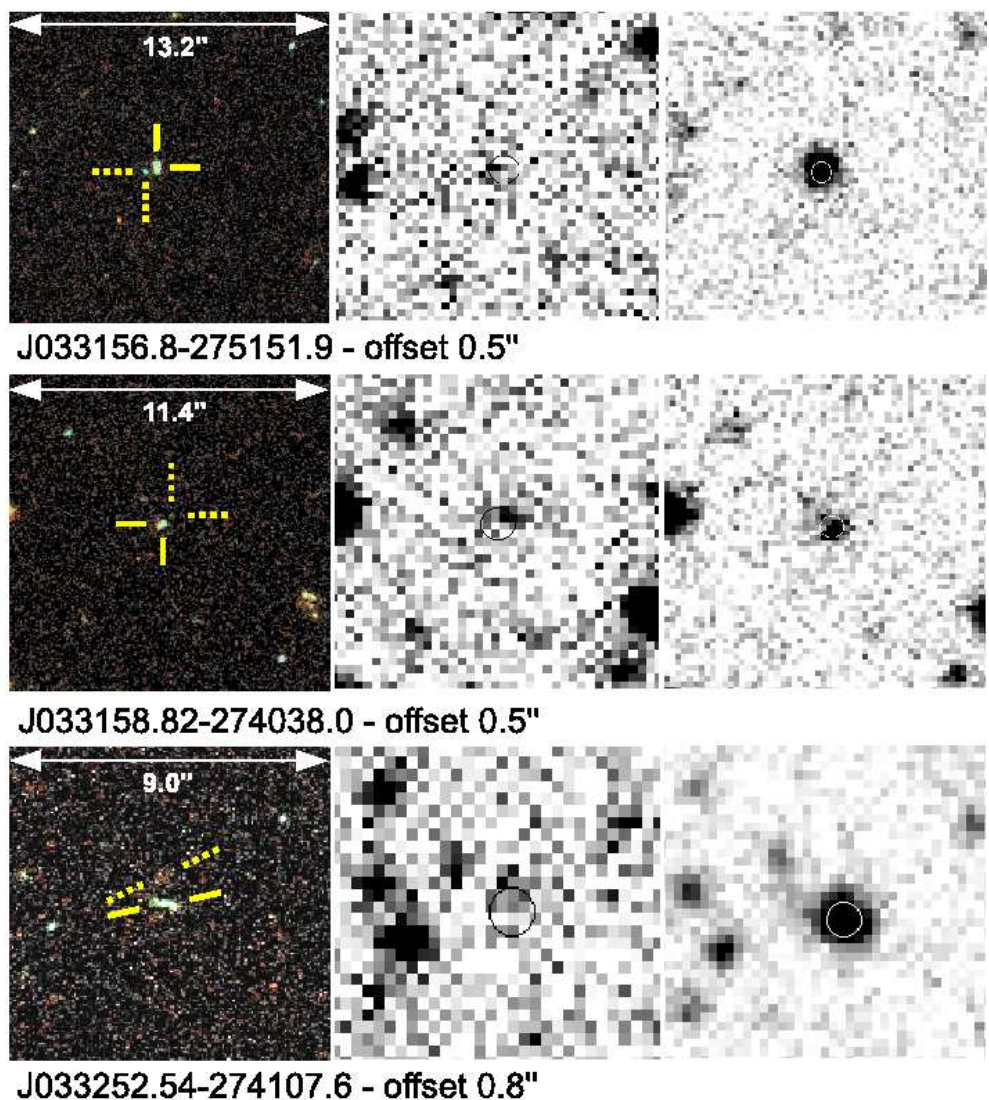} 
\caption{Images of the three $z>3.4$ LBGs outside the ACS GOODS-South area  with
offset IB detections (S/N$\sim$2.5). From left to right, ACS two-color images (from GEMS), IB and VIMOS $R$-band images are
shown. 
The dotted and solid lines indicate the possible nearby polluting source and the targeted
LBG, respectively. 
In all three cases the offset emission in the IB is consistent with the presence of a close source
visible in the ACS images. 
In the middle $U$-band images, black circles outline the 1.2\see~diameter apertures, while in the
$R$-band images white circles indicate the position of the spectroscopic target. Below the images, to the right of the GOODS ID,
the separation between the LBG and the nearby source is reported.
The spectrum of the top source is presented in Figure~\ref{SPEC2D}.
  \label{extended}}
\end{figure}

\begin{figure}
 \epsscale{1.0}
 \plotone{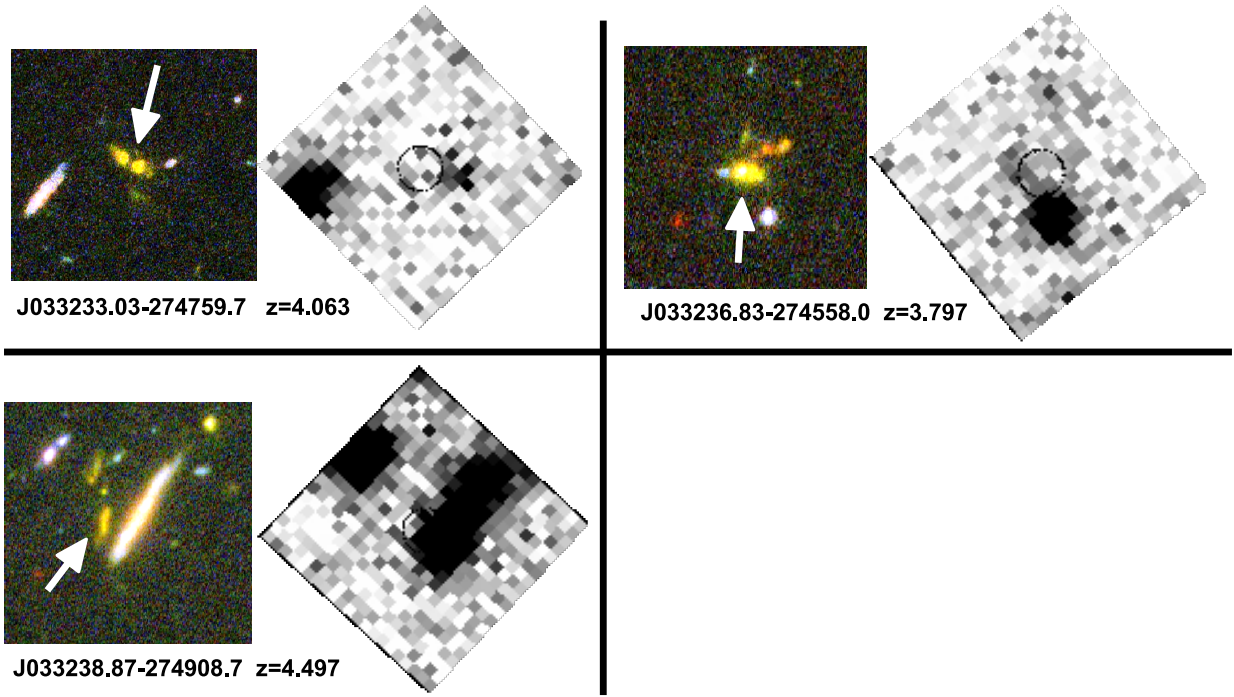} 
\caption{Three LBGs in the HUDF detected at S/N$>$1 (1.2\see~diameter) in the IB image. 
The $BVi$ color images at the HUDF depth are shown on the left of 
each panel.
The position of the LBG is marked with a solid arrow in the color image and a circle in the IB (black/white)
image. Blue compact sources detected in the IB images are visible, both close to the LBG and in the field.
The box size of the IB cutouts is 6\see on a side. \label{Fig4}}
\end{figure}

\begin{figure}
 \epsscale{1.}
 \plotone{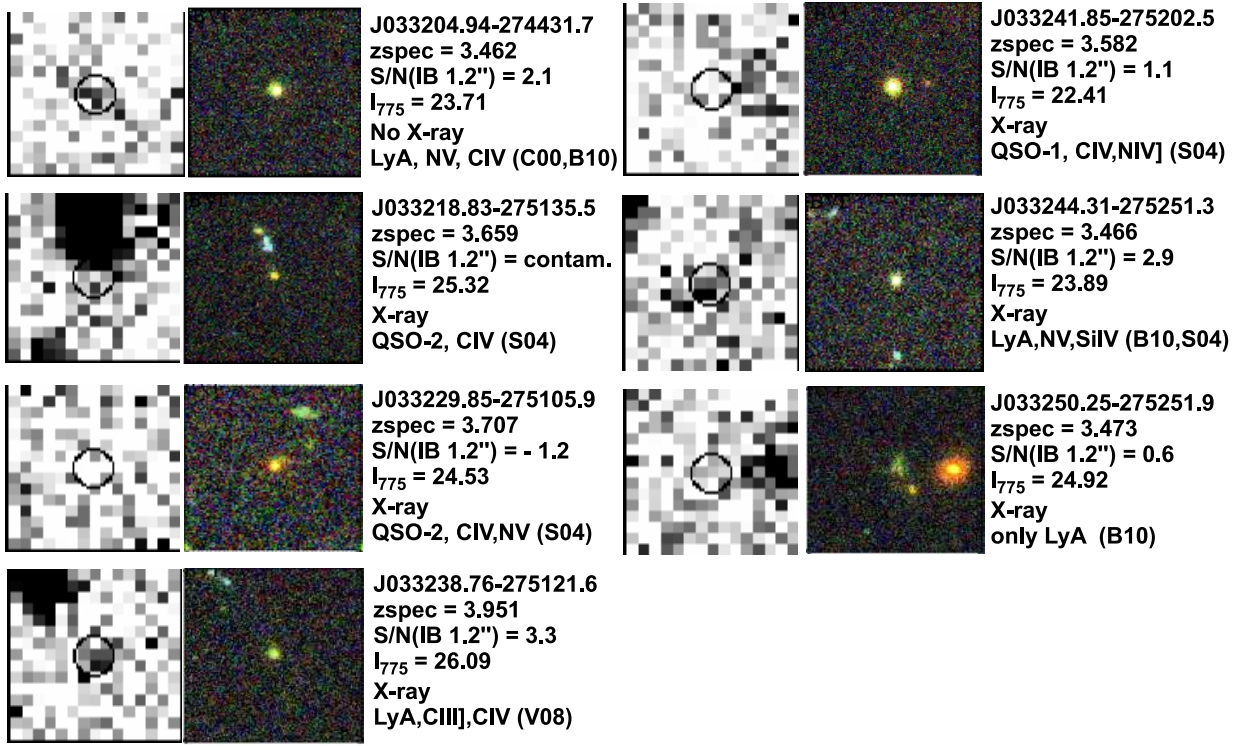} 
\caption{{\it HST}/ACS $BVi$ color images and the ultra-deep VLT/VIMOS IB cutouts 
for the seven AGNs with spectroscopic redshift higher than 3.4. The circles in the IB images have 1.2\see~diameters and the box sizes are 4.5\see on a side. For each pair the GOODS~ID, redshift, S/N ratio within the 1.2\see~diameter aperture, \wi\ magnitude,
and the information on the X-ray detection and spectral properties are reported. C00, B10, S04, V08 correspond to
Cristiani et al. (2000), Balestra et al. (2010), Szokoly et al. (2004) and Vanzella et al. (2008), respectively.
Three out of seven AGNs show a LyC detection at S/N$>$2.  \label{AGNs}}
\end{figure}

\begin{figure}
 \epsscale{1.0}
 \plotone{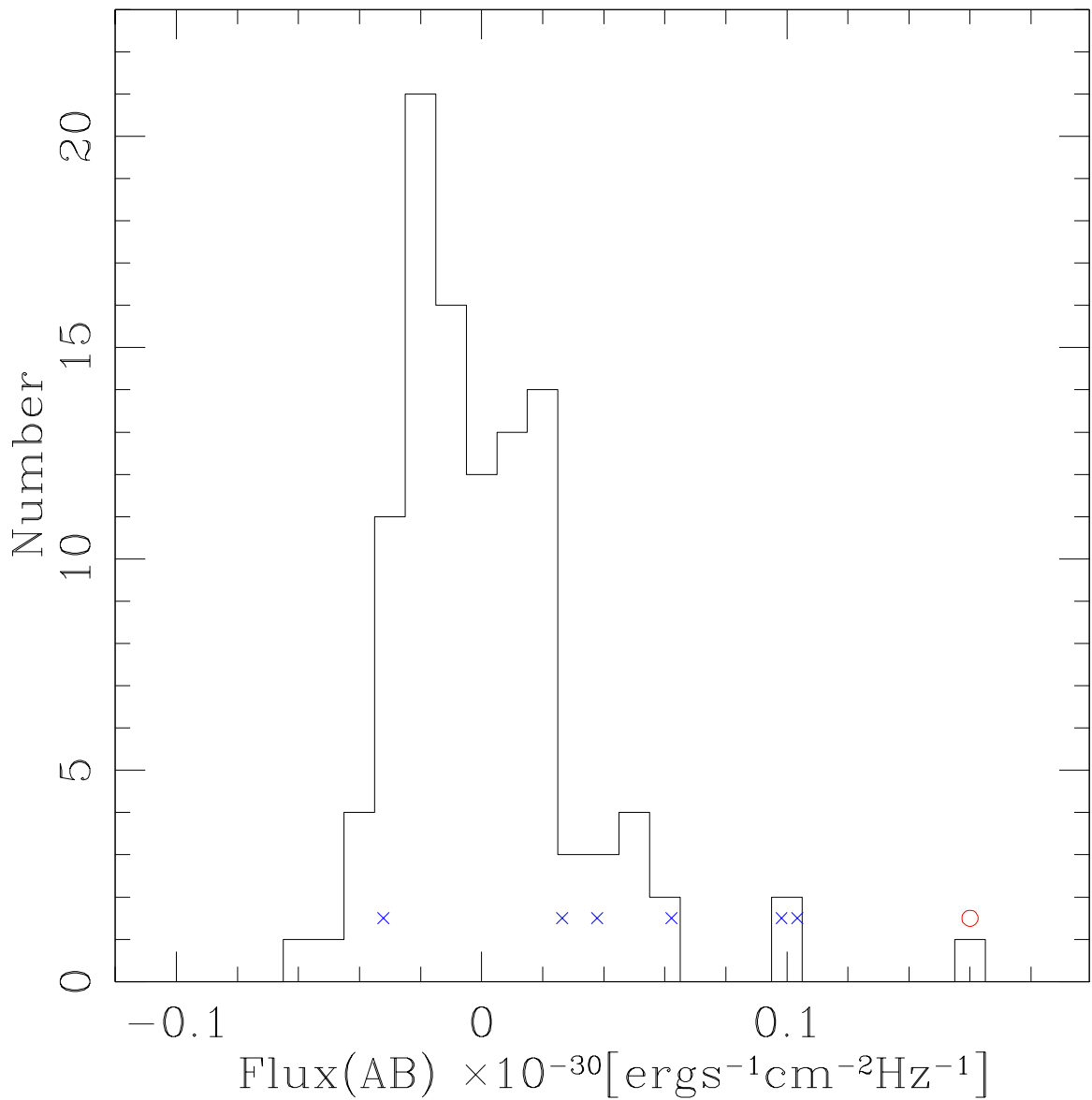} 
\caption{Flux distribution of the clean LBG sample (102) and AGNs (6) in AB units and 
within 1.2\see~diameter aperture is shown. AGNs are marked with blue crosses, and the LBG with 
a red circle. Three AGNs (from right to left) and the LBG (circle) have been detected 
in their LyC with S/N higher than 2 (see also Table~\ref{tab:sampleIB}).
\label{Fluxes}}
\end{figure}

\begin{figure}
 \epsscale{0.95}
 \plotone{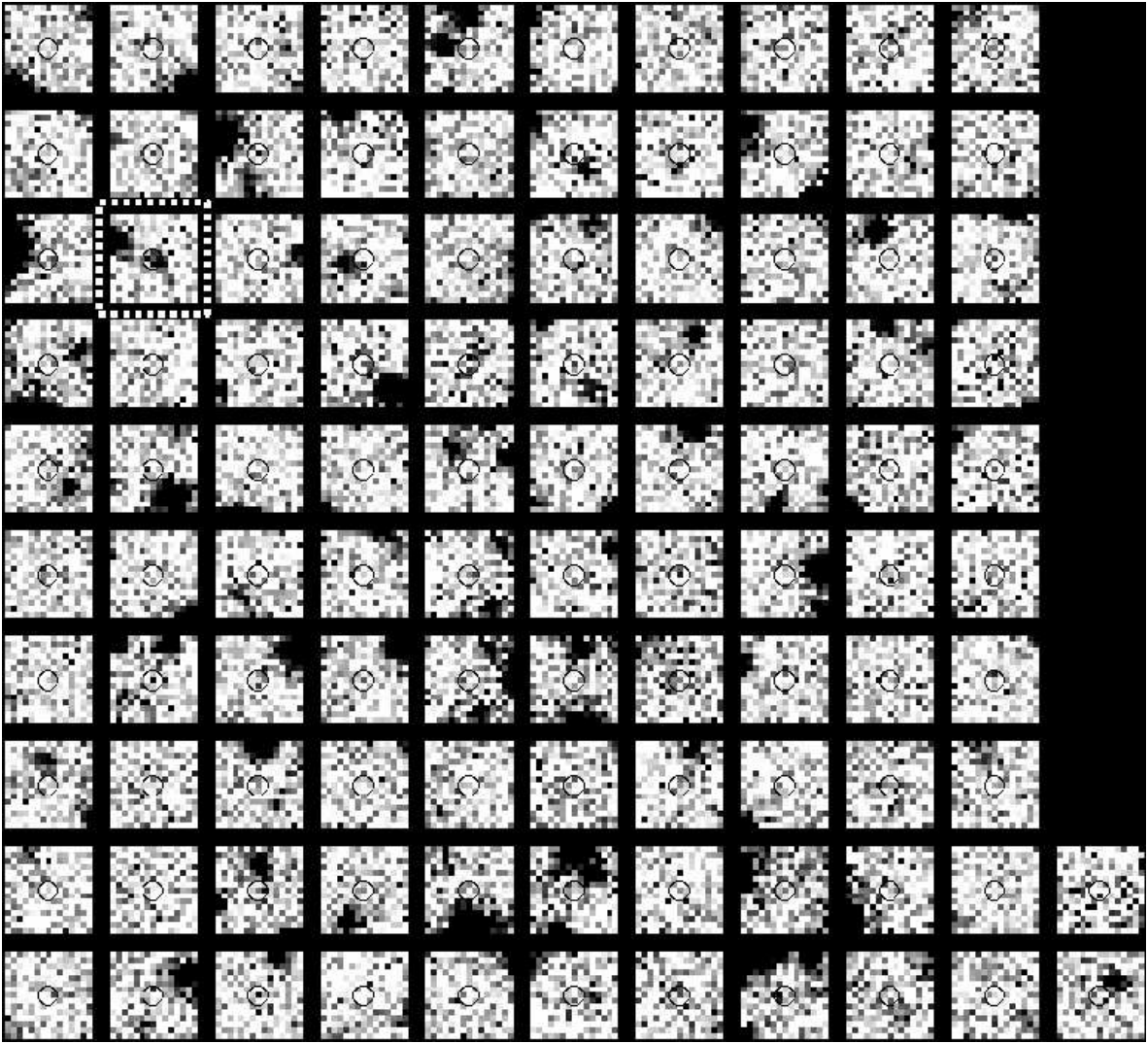} 
\caption{VLT/VIMOS $U$-band cutouts of the sources adopted in the simulations (clean sample). The box sizes are 4.5\see~ on a side.
The sole LBG detected in LyC with S/N higher than 2 in the 1.2\see~diameter aperture is marked with a dotted square 
(GDS~J033216.64-274253.3 with S/N=5.5, described in Sect. 3.2.1). Circles indicate the position of the 1.2\see~diameter apertures. \label{CLEANpano}}
\end{figure}

\begin{figure}
 \epsscale{0.28}
 \plotone{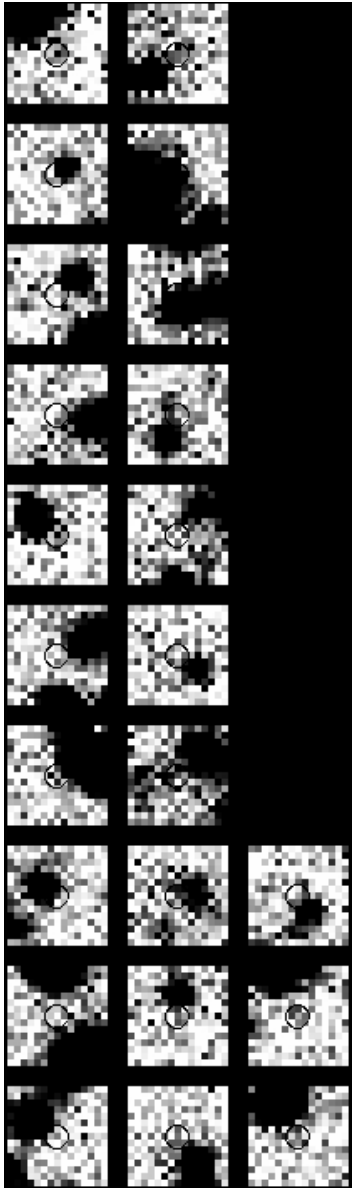} 
\caption{VLT/VIMOS $U$-band cutouts of the sources excluded from the simulations because of the presence of a nearby 
blue object. The box sizes are 4.5\see~ on a side (see text for details).\label{BADpano}}
\end{figure}

\begin{figure}
 \epsscale{0.9}
 \plotone{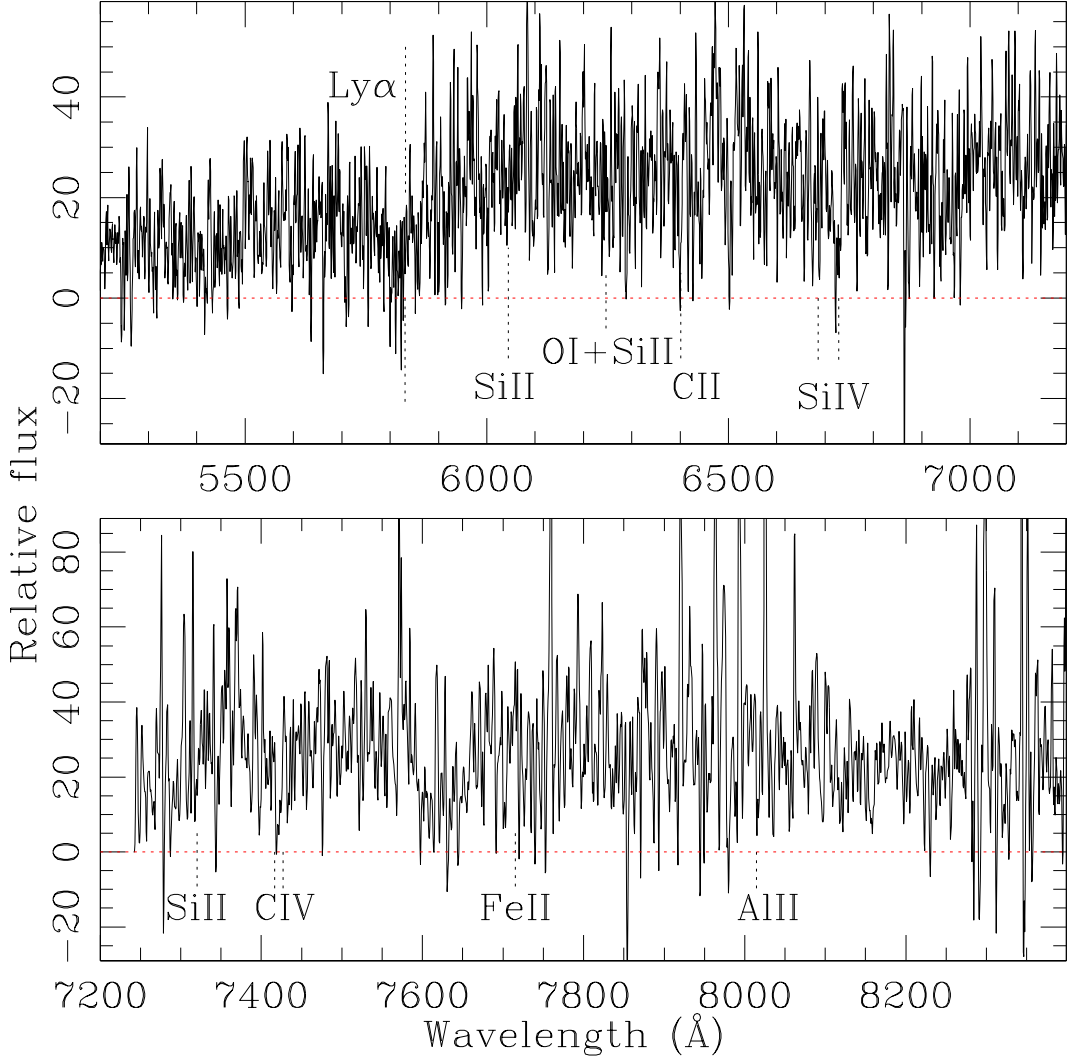} 
\caption{Extracted Keck-DEIMOS spectrum of the LBG GDS~J033216.64-274253.3 with LyC detection in the IB image (S/N$\simeq$5.5).
In the top and bottom panels the blue and red parts of the spectrum are shown; \Lya, SiIV 1403\AA~ and CIV 1548-1550\AA~absorptions
are clearly seen, and we marginally detect CII 1335\AA~absorption. 
Absorption from SiII 1260, OI+SiII 1302-1304, SiII 1526, FeII 1608 and AlII 1671 are not detected, nor
are emission lines like NiV] 1486 and HeII 1640 detected.
A comparison with the cB58 spectrum with the IRAF task $rvsao$ gives 
a good cross correlation coefficient (R=3.34) and a redshift of 3.797. \label{LBGdetect}}
\end{figure}

\begin{figure}
 \epsscale{1.0}
 \plotone{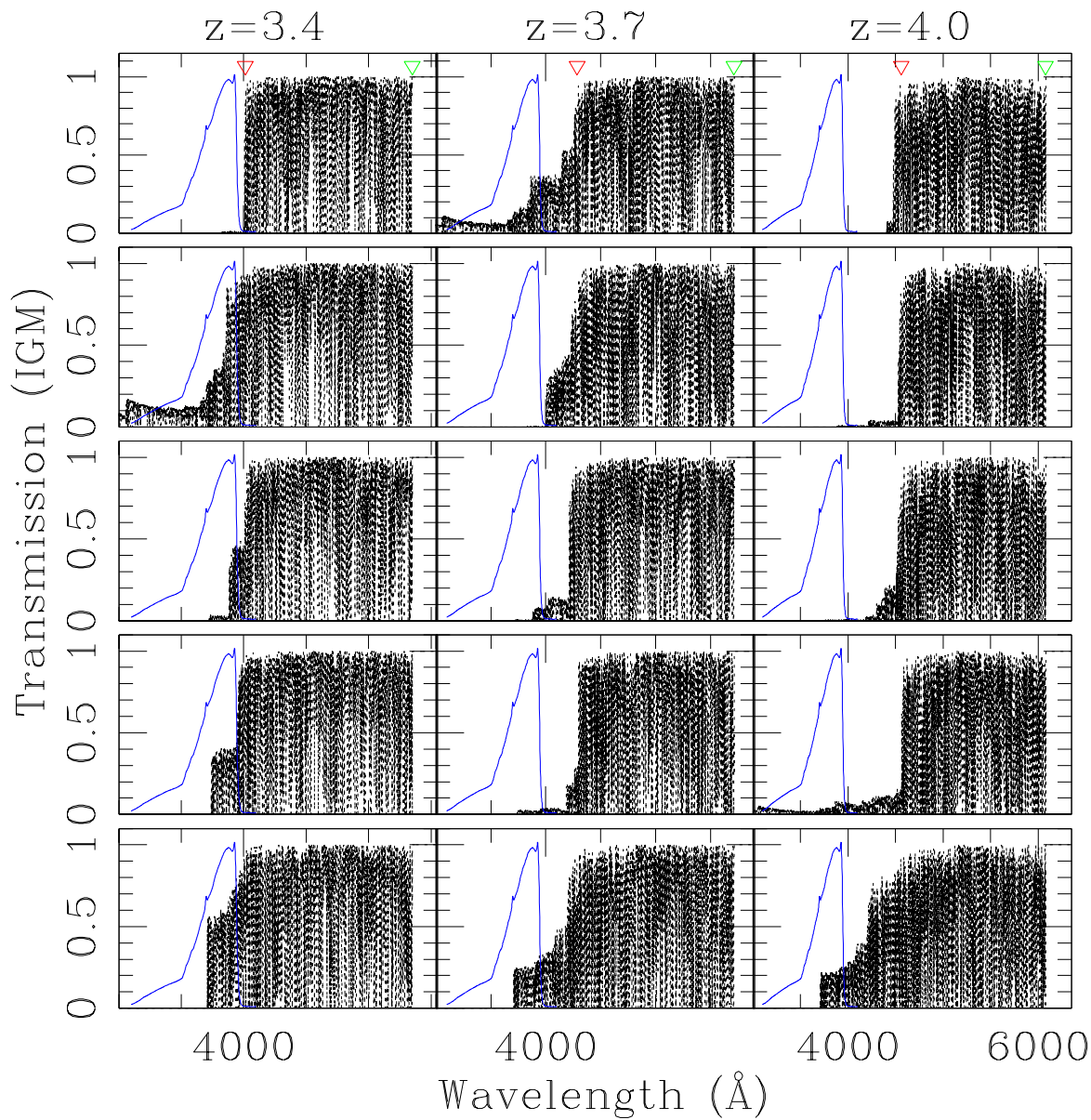} 
\caption{Examples of IGM transmission derived from the IW08 models for three redshift values.
 Zero transmission at blue wavelengths occurs when there is a LLS or DLA system at lower redshift.
 The IB filter shape is shown as dotted blue lines and in the top panels the positions of the 912\AA~Lyman limit
and \Lya\ are marked with open red (left) and green (right) triangles, respectively. 
  \label{TransSingle}}
\end{figure}

\begin{figure}
 \epsscale{1.0}
 \plotone{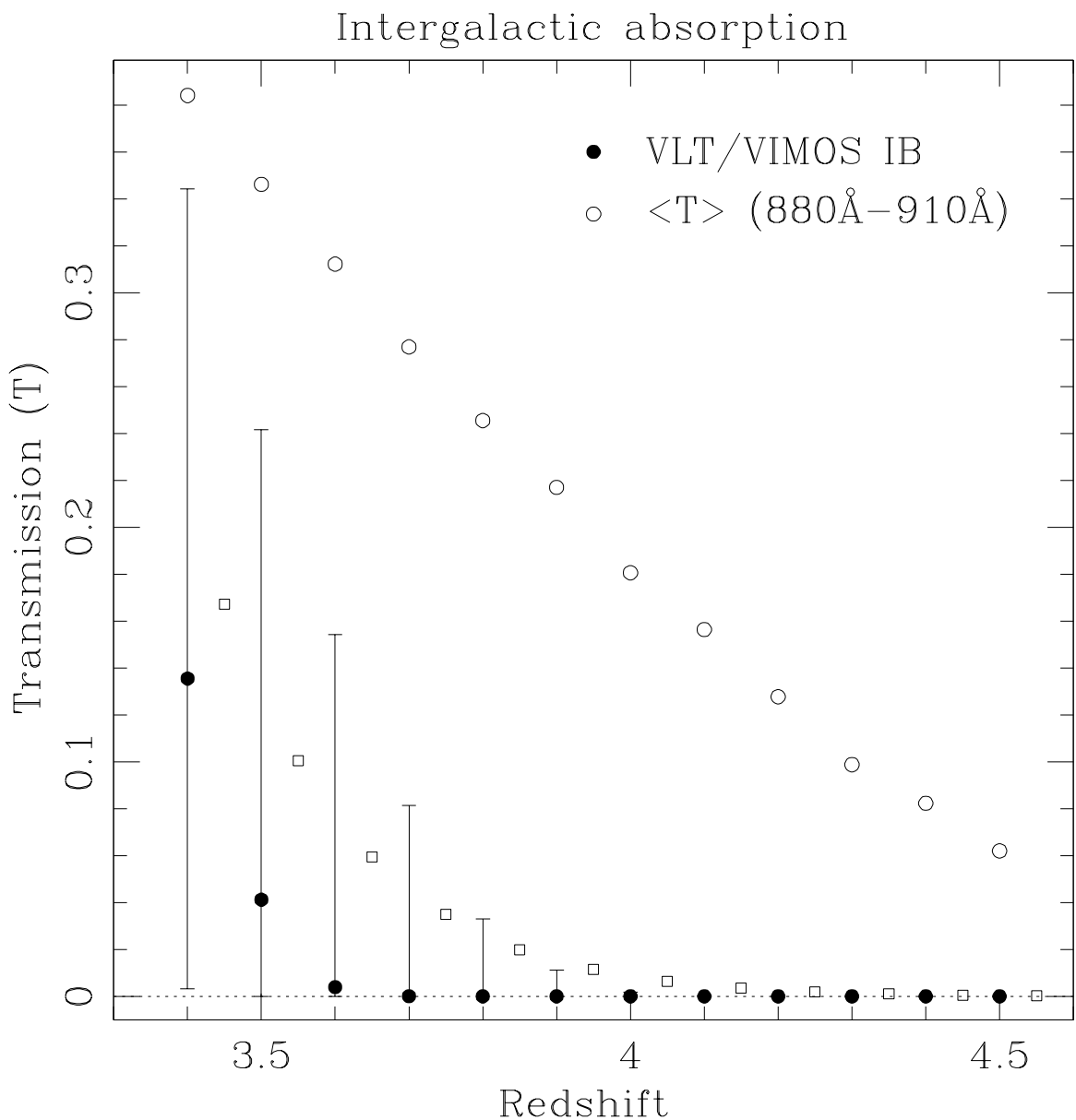} 
\caption{Transmission averaged over the wavelength range of the LyC (880-910 \AA) (open circles) and convolved 
 with the VLT/VIMOS IB filter (filled circles) as a function of source redshift. The filled circles and vertical error bars
 indicate the median value and central 68\% range of the transmission for the 10,000 lines of sight generated with the IW08 simulations. Open squares 
 are the averages calculated over the same lines of sight (shifted by $dz$=0.05 to the right for clarity).  Clearly
 the VLT/VIMOS IB probes progressively shorter wavelengths as redshift increases, with the effect of lowering the transmission.
  \label{Trans}}
\end{figure}

\begin{figure}
 \epsscale{1.0}
 \plotone{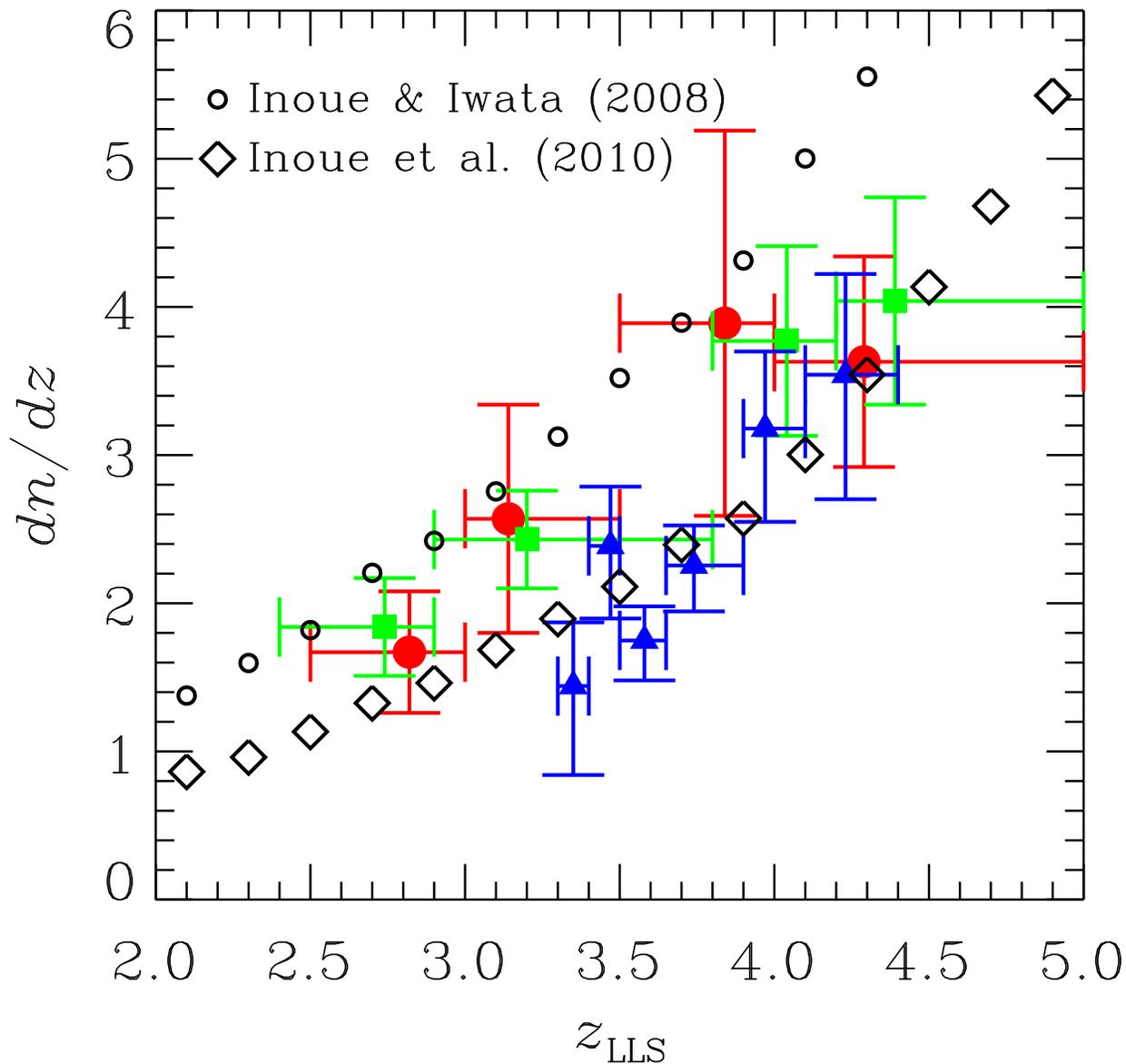} 
\caption{Number density of LLSs vs. redshift reported from previous and recent works. 
The filled circles are estimations of P\'eroux et al. (2005), squares are Songaila \& 
Cowie (2010),
and triangles are Prochaska et al. (2010). Open circles are the number density adopted in IW08,
while diamonds are that assumed in the  updated simulation by Inoue et al. (2010).
  \label{new_lls}}
\end{figure}

\begin{figure}
 \epsscale{1.0}
 \plotone{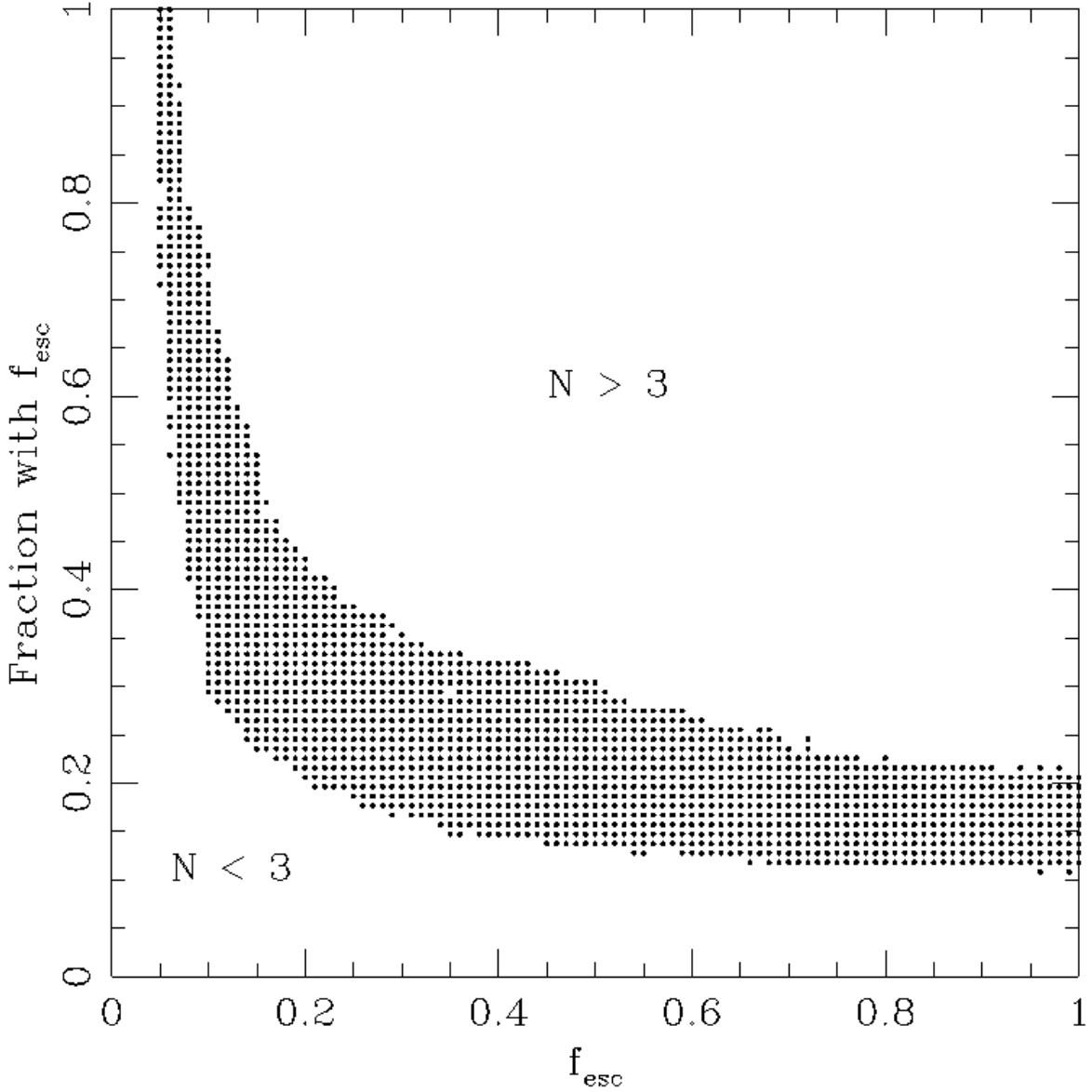} 
\caption{Parameter space excluded by a Monte Carlo analysis described in the text.
The x-axis is the escape fraction ($f_\mathrm{esc}$),
and the y-axis is the fraction of galaxies that have this escape fraction. The other
galaxies are assumed to have negligible escape fractions. The shaded region correspond to
a number of expected LyC detections in the IB survey equal to 3. The fact that only 1 out of 
102 LBGs has been detected suggest that our observations are compatible with the 
region where N$<$3, implying that large fractions of the sample with high 
$f_\mathrm{esc}$ are excluded (N$>$3).  \label{check}}
\end{figure}

\begin{figure}
 \epsscale{1.0}
 \plotone{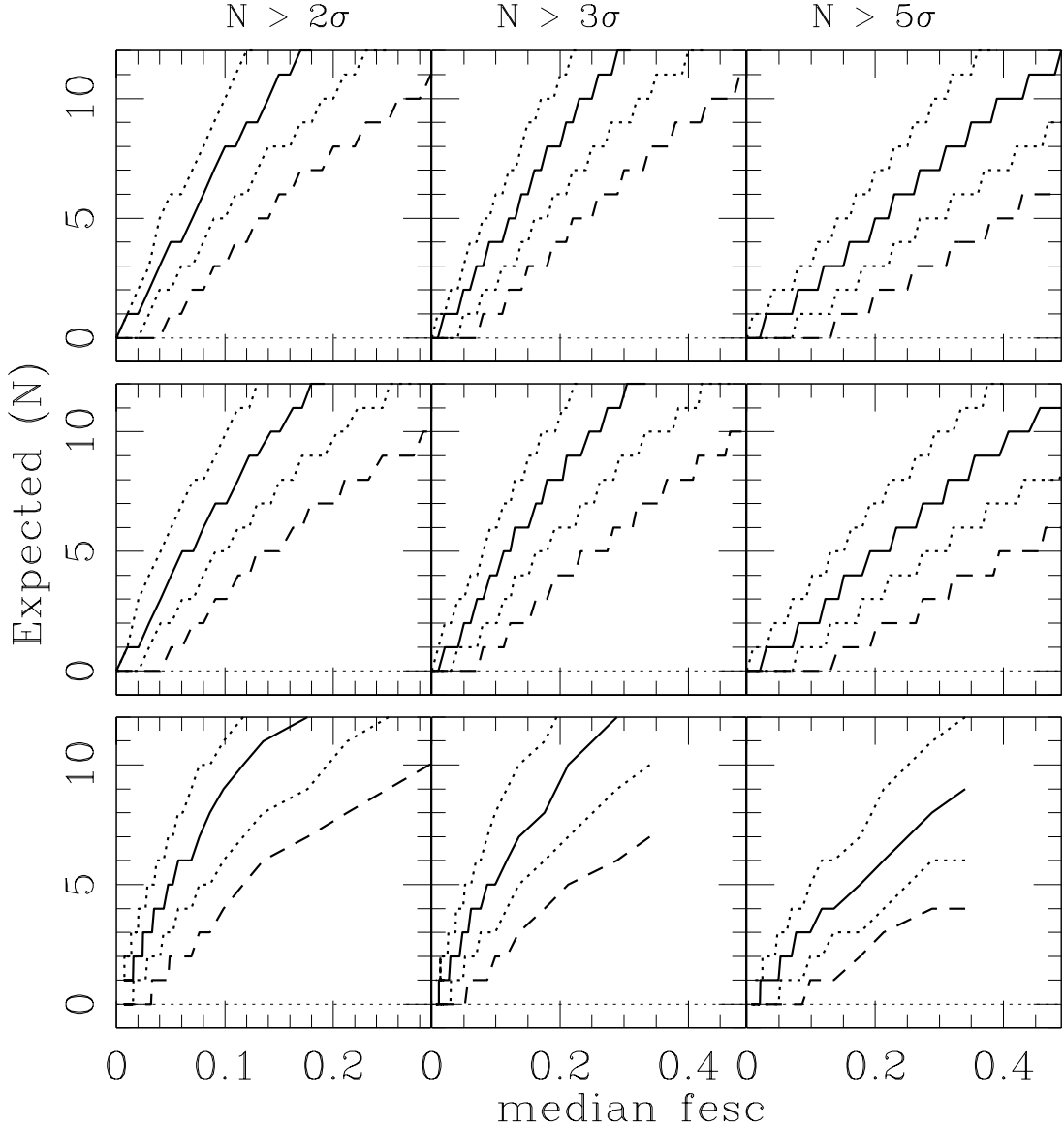} 
\caption{Monte Carlo simulations of the expected number of LyC detections in the ultra-deep VLT/VIMOS $U$-band imaging
 as a function of the median $f_\mathrm{esc}$ (in the left panels $f_\mathrm{esc}$ up to 30\% is shown,
 in the middle and right panels it is shown up to 50\%).
 From left to right, the expected number of LyC detections are presented for three IB depths, 2$\sigma$ (29.5), 
 3$\sigma$ (29.1) and 5$\sigma$ (28.6), respectively.
 Solid lines show the median expected number, while the dotted lines and dashed line mark the one sigma and
 three sigma limits, respectively.
 In the top panels a constant $f_\mathrm{esc}$ value is assumed, from 1\% to 100\%. 
 In the middle and bottom panels Gaussian and exponential distributions of $f_\mathrm{esc}$ with different
 medians are shown, respectively.  The abscissa reports the median of the simulated distribution 
 (see text for details).
 \label{MC}}
\end{figure}

\begin{figure}
 \epsscale{0.85}
 \plotone{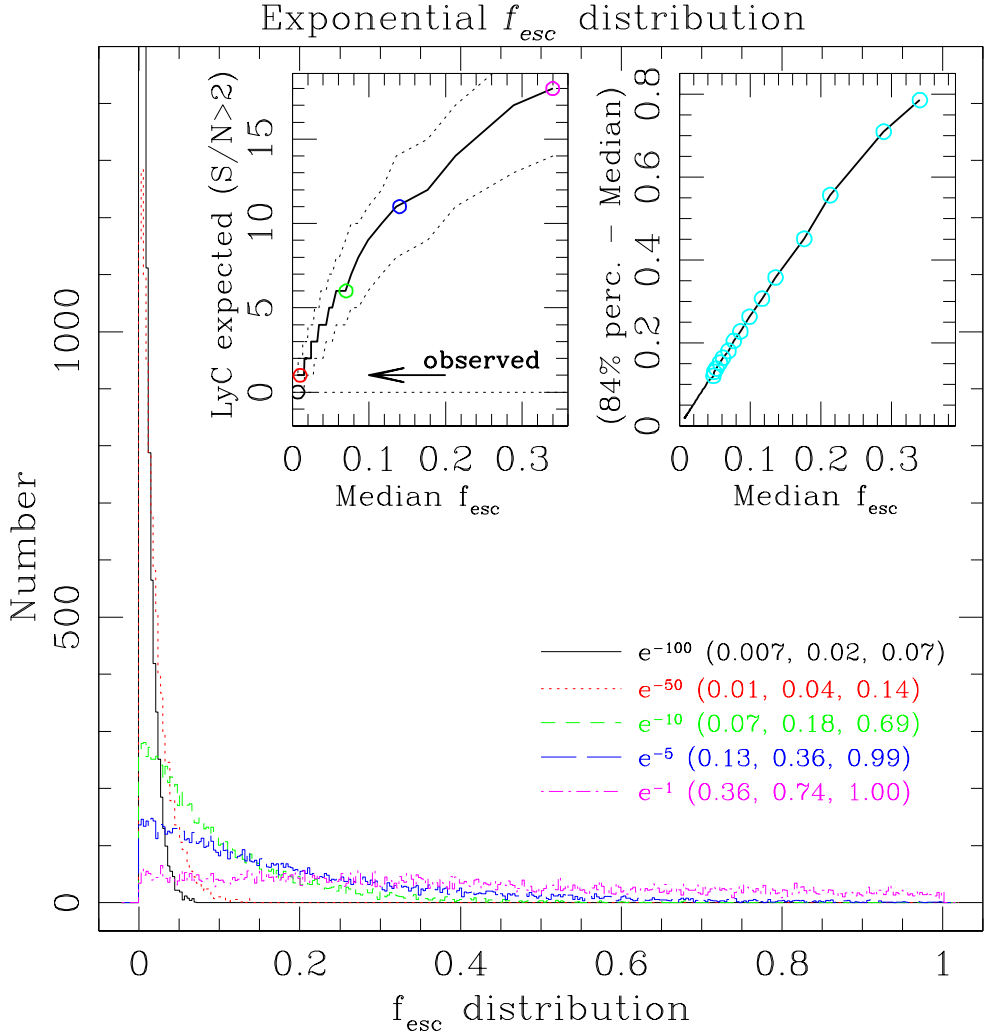} 
\caption{
{\bf Left inner box:} Monte Carlo simulations of the expected median number of LyC detections 
in the ultra-deep VLT/VIMOS $U$-band imaging (thick solid line) and 68 percent central 
interval (dotted lines) reported as a function of the median of the 100 exponential distributions 
explored (different slopes $\lambda$ have been explored). 
Color-coded open circles correspond to the examples of distributions shown in the main box.
{\bf Right inner box:} The black solid line outlines the region occupied 
 by the 100 distributions in the plane (1$\sigma$, median). The distributions for which the expected
number of LyC detection is larger or equal to 5 (probability less than 5\% to observe $\leq$ 1 LyC detection)
are shown with open cyan circles. 
{\bf Main box:} Examples of exponential distributions of $f_\mathrm{esc}$ with $\lambda$=1, 5, 10, 50, 100 are shown 
(10,000 extractions for each one have been done). The numbers reported in the legend 
from left to right are the median, the 84\% percentile and the maximum value, respectively. 
\label{MC_exp}}
\end{figure}

\begin{figure}
 \epsscale{0.85}
 \plotone{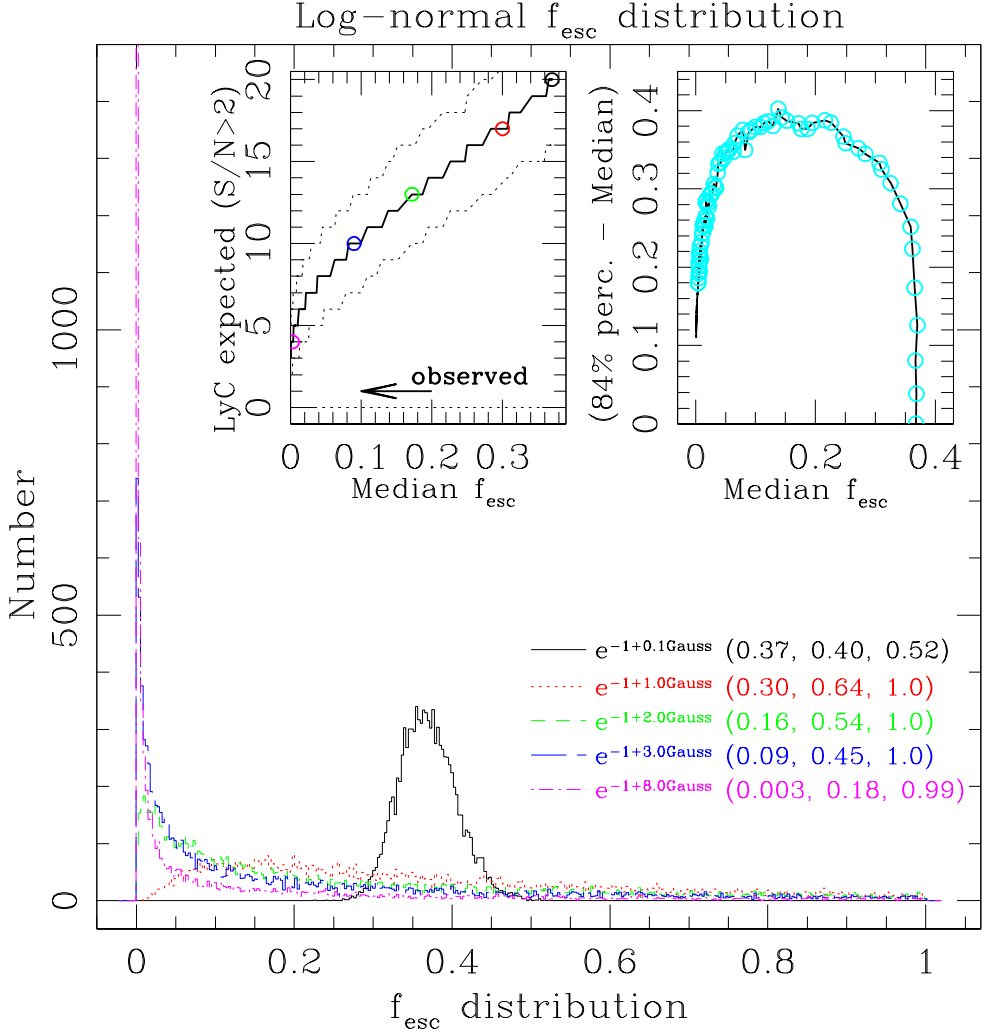} 
\caption{
 {\bf Left inner box:} Monte Carlo simulations of the expected median number of LyC detections
 in the ultra-deep VLT/VIMOS $U$-band imaging (thick solid line) and 68 percent central 
 interval (dotted lines) reported as a function of the median of the 100 log-normal distributions 
 explored (the case with K=1 and $\lambda$ running from 0.1 to 10 with step 0.1 is shown). 
 Color-coded open circles correspond to the examples of distributions shown in the main box. 
 {\bf Right inner box:} The black solid line outlines the region occupied
 by the 100 distributions in the plane (1$\sigma$, median). The distributions for which the expected
 number of LyC detection is larger or equal to 5 (probability less than 5\% to observe $\leq$ 1 
 LyC detection) are shown with open cyan circles. 
 {\bf Main box:} Examples of log-normal distributions of $f_\mathrm{esc}$ with $\lambda$=0.1, 1.0, 2.0, 3.0 
 and 8.0 are shown (10,000 extractions for each one have been done). The numbers reported in the legend
 from left to right are the median, the 84\% percentile and the maximum value, respectively.
 \label{MC_LOGN}}
\end{figure}

\begin{figure}
 \epsscale{0.9}
 \plotone{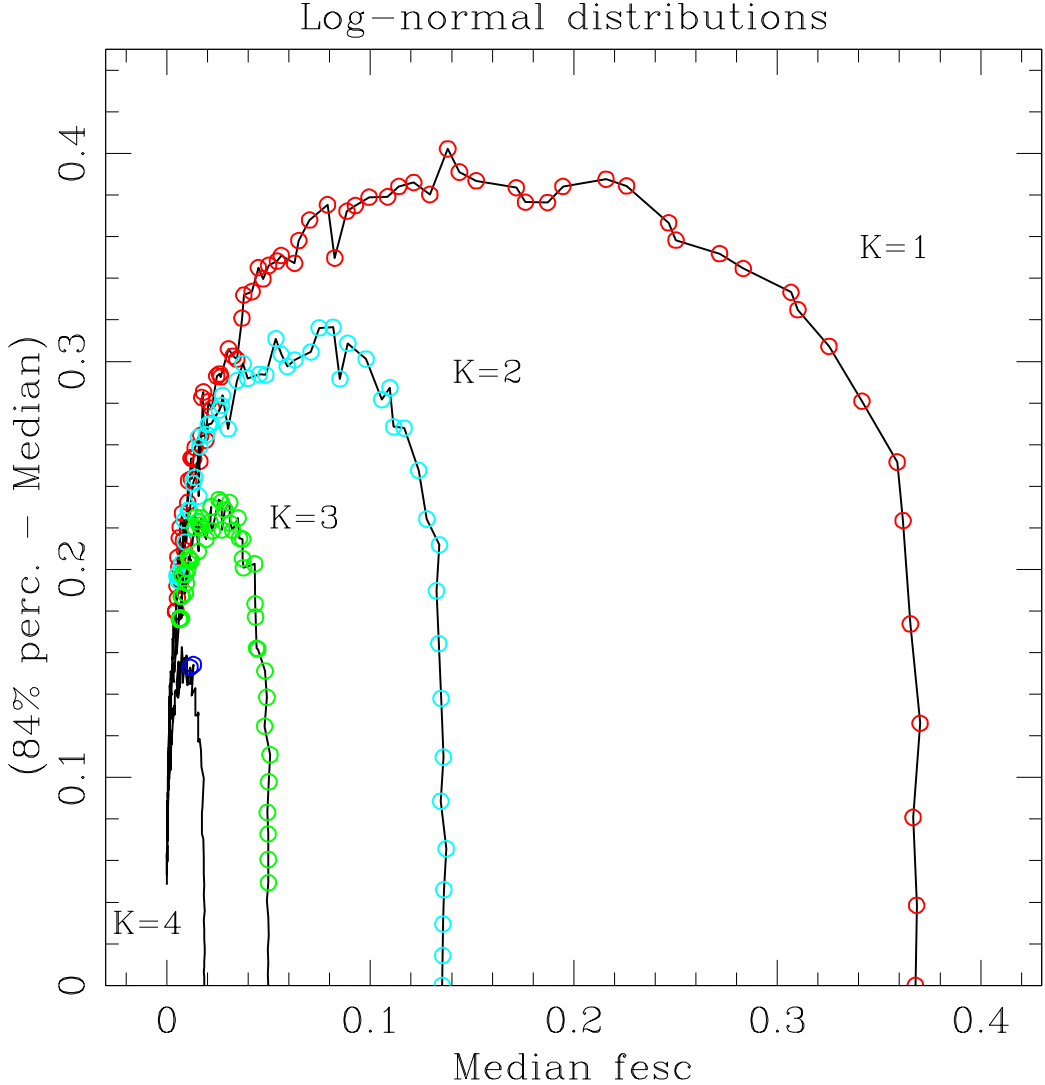} 
\caption{The same as shown in the inner right box of Figure~\ref{MC_LOGN}, but calculated
for different $K$ values of the log-normal distributions ($e^{-K+\lambda \times Gauss}$).
The black solid lines show the regions occupied by the 100 distributions for each $K$ value.
The distributions for which the expected
number of LyC detection is larger or equal to 5 are shown with open circles.
The single observed LyC detection reported in the present work suggests that  $f_\mathrm{esc}$ 
is distributed with median and 1$\sigma$ upper tail lower than $\sim$ 6\% and 18\%, 
respectively, if a log-normal distribution is assumed. \label{MC_LOGN_MED_SIG}}
\end{figure}

\begin{figure}
 \epsscale{1.0}
 \plotone{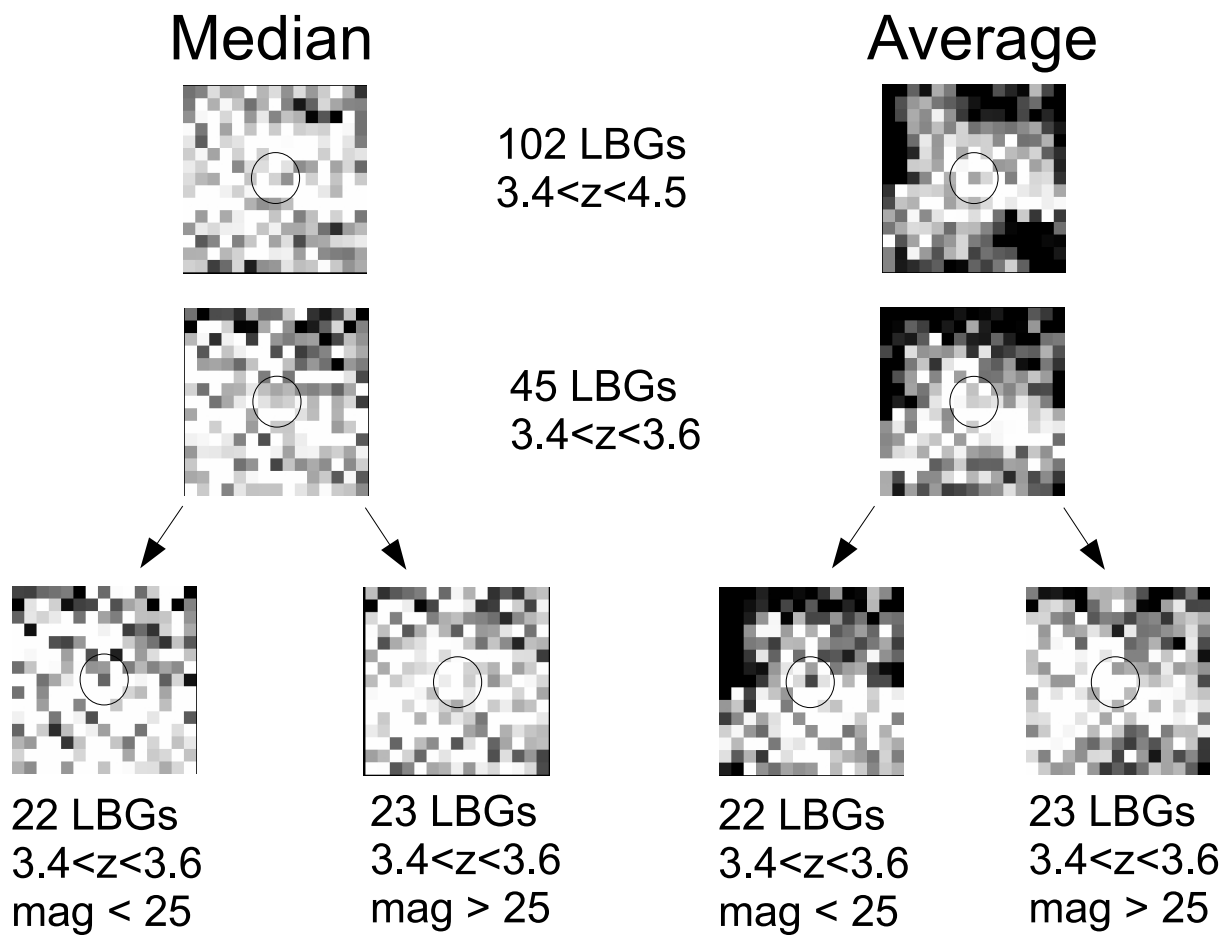} 
\caption{Median stacking of LBGs from the $clean$ spectroscopic sample. Stacking of redshift-selected sub-samples has been
performed in order to increase the IGM transmission, as well as for the brighter (\wi\ $<$ 25) and 
fainter (\wi $>$ 25) sub-samples (see text). The pixel size is 0.3\see~and each box is 4.5\see~ on a side. The circles
have diameters of 1.2\see.
   \label{STACK}}
\end{figure}

\begin{figure}
 \epsscale{0.3}
 \plotone{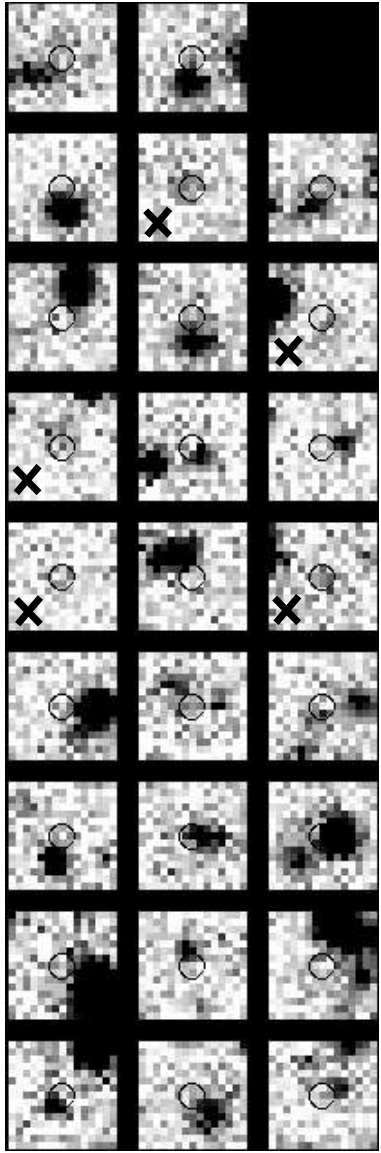}
\caption{VLT/VIMOS $U$-band cutouts of the sources selected in the HUDF with magnitude 27 $<$ \wi\ $<$ 28.5, photometric 
redshifts in the range $3.4 < z_{\rm phot} <4.0$
and S/N ratio in the 1.2\see~diameter aperture higher than 2. The size of the boxes is 4.5\see~on a side. 
Sources indicated with a black cross show a non-offset detection in the IB.\label{CLEANpanoBPZ}}
\end{figure}

\begin{figure}
 \epsscale{1.0}
 \plotone{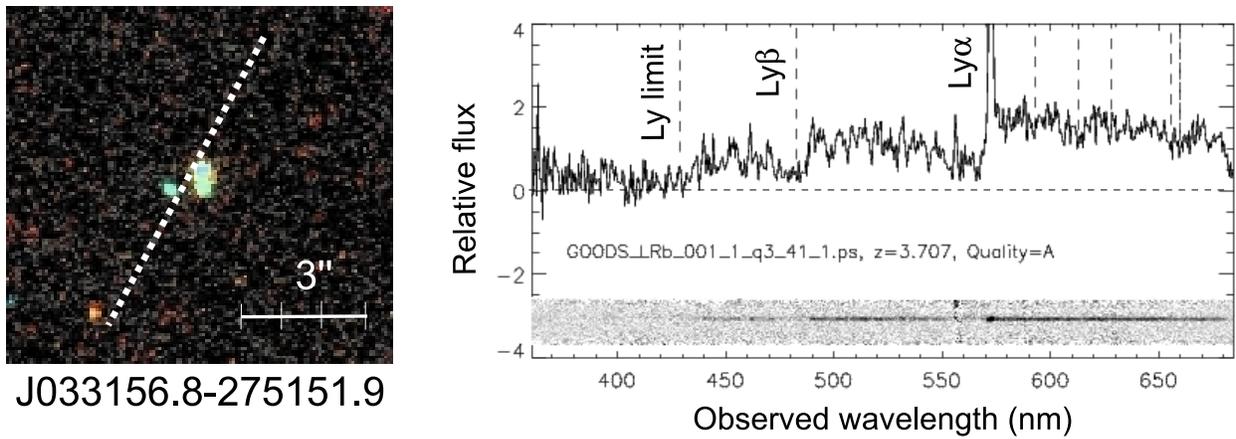}
\caption{Two-dimensional spectrum of the source GDS 033156.8-275151.9 in the outer region of the GOODS-South.
On the left side the {\it HST}/ACS color image derived from the GEMS survey; the dotted line shows the slit orientation
in the sky. As discussed in the text, a compact source close to the LBG (center) at sub-arcsecond separation is clearly
visible (to the left). The spectrum (right part of the figure) contains the contribution of both sources. In particular,
faint flux is detected below the Lyman limit, most probably due to the close (lower redshift) companion.  \label{SPEC2D}}
\end{figure}

The ESO/VIMOS spectroscopic survey extends beyond the deep GOODS-South area,
but where the IB photometry is still avaliable. Thirteen galaxies with secure redshifts match 
the selection 3.4$<z<$4.5.
Three out of 13 show an IB detection with S/N$\simeq$2.5. 
In all three cases the IB emission is offset with respect to the position of the LBG. High--resolution 
imaging ({\it HST}/ACS) drawn from
the Galaxy Evolution from Morphology and SEDs project (GEMS, Rix et al. 2004) have been used to check for the
presence of close companions that may contaminate the IB photometry. Even though the GEMS survey
is shallower than GOODS, in all cases there is a distinct faint source shifted in the direction
consistent with the IB emission
(Figure~\ref{extended} shows the {\it HST}/ACS color (\wv\ and \wz\ combined), the VIMOS IB and 
$R$ images, where the deep $R$ data are described in Nonino et al. 2009). 
In particular for the source J033156.8-275151.9 (top panel), a distinct compact source 
at $\sim$ 0.6\see~separation from the LBG is clearly present 
(marked with dotted lines in the Figure). In this case a signal bluer than the Lyman limit is also visible 
in the two dimensional spectrum (see Figure~\ref{SPEC2D}). There is no spatial offset between the two traces in the 
spectrum because of the slit orientation over the sky superposes the two
objects along the wavelength dispersion. It is further confused by the seeing 
conditions during the observations ($\sim$ 1\see). No additional spectral features, possibly
arising from the close source, have been detected.
In the other two cases (middle and bottom panels of Figure~\ref{extended}) a similarly offset
and faint close source is present that can be linked to the contamination.

\clearpage
\begin{deluxetable}{lcc}
\tabletypesize{\scriptsize}
\tablecaption{Aperture magnitude limits of the IB VIMOS and NB FORS1 surveys in the GOODS-South. \label{tab:mags}}
\tablewidth{0pt}
\tablehead{
\colhead{Depth} & \colhead{$U$-VIMOS} & \colhead{NB-FORS1 }
}
\startdata
5$\sigma$       &   28.6            &     27.1          \\  
1$\sigma$       &   30.5            &     29.0          \\
\tableline
\enddata
\tablecomments{Magnitudes are reported within aperture diameters of 1.2\see.}
\end{deluxetable}

\clearpage
\begin{deluxetable}{lccc}
\tabletypesize{\scriptsize}
\tablecaption{Summary of the sources adopted. \label{tab:spec}}
\tablewidth{0pt}
\tablehead{
\colhead{N} & \colhead{AGNs [N/Detect/Cont]} & \colhead{LBG [N/Detect/Cont]} & \colhead{isolated LBG}
}
\startdata
122 GOODS-South              &   7/3/1           &     115/1/23    & 92   \\
13  Ext GOODS-South          &   0/0/0           &     13/0/3      & 10     \\
\hline
135 Total                &   7/3/1           &     128/1/26    &   {\bf 102}  \\
\tableline
\enddata
\tablecomments{[N/Detect/Cont] indicates the number of sources (N), the number LyC detections (Detect)
and the number affected by nearby sources (Cont). ``Ext GOODS-South'' indicates the extended GOODS-South region.}
\end{deluxetable}

\clearpage
\begin{deluxetable}{lccccccl}
\tabletypesize{\scriptsize}
\tablecaption{The spectroscopic sample of galaxies with a detection
  in the IB image with S/N$>$2 in 1.2\see~diameter aperture.   \label{tab:sampleIB}}
\tablewidth{0pt}
\tablehead{
\colhead{GOODS ID} & \colhead{S/N 1.2\see} & \colhead{S/N 2.1\see} &   
\colhead{\wi\ } & \colhead{\wi\ - \wz\ } & \colhead{zspec} & 
\colhead{$\beta$} & \colhead{comment}
}
\startdata
        J033204.94-274431.7    &2.1 & 7.2  &23.71& 0.019&3.462\tablenotemark{d,b}&-1.9&  AGN, NV, CIV; X-ray No\\ 
        J033216.64-274253.3    &5.2 &4.4  &24.86&-0.015& 3.795\tablenotemark{c}&-2.1&  LBG, SiIV, CIV (abs); X-ray No\\ 
        J033238.76-275121.6        &  3.3    & 3.4 &26.09& 0.11&3.951\tablenotemark{f,a} &-1.5 & AGN, CII,CII[,CIV; X-ray Yes\\   
        J033244.31-275251.3    & 2.9& 2.0 &23.89&-0.11&3.466\tablenotemark{c,e}&-2.6&  AGN, NV, SiIV; X-ray Yes\\ 

\tableline
\enddata
\tablenotetext{a}{From Vanzella 08, 09.}
\tablenotetext{b}{GOODS VIMOS/LRB, P08, B10.}
\tablenotetext{c}{Stern et al. in preparation.}
\tablenotetext{d}{Cristiani et al. 2000, A\&A, 355, 485, multiple zspec.}
\tablenotetext{e}{Szokoly et al. 2004, ApJS, 155, 271.}
\tablenotetext{f}{Spectrum re-analyzed, new redshift measure: CIV, CIII, \Lya.}
\end{deluxetable}

\clearpage
\begin{deluxetable}{lcccccc}
\tabletypesize{\scriptsize}
\tablecaption{Upper limits on $f_\mathrm{esc,rel}$ from stacking.  \label{tab:fesc}}
\tablewidth{0pt}
\tablehead{
\colhead{\wi$<$24.75} & \colhead{\wi$<$25.0} & \colhead{\wi$<$25.25}& \colhead{\wi$<$25.5}& \colhead{\wi$<$25.75}&\colhead{redshift}&\colhead{$<T>$}
}
\startdata
                 0.051(10)    &0.047(17)    &0.045(26)    &0.045(28)    &0.046(30)     & [3.40--3.55] &0.135\\ 
                 0.136(17)    &0.130(21)    &0.127(38)    &0.127(36)    &0.127(37)     & [3.55--3.75] &0.050\\  
                 0.058(27)    &0.054(38)    &0.052(56)    &0.053(64)    &0.053(67)     & [3.40--3.75] &0.090\\ 
                 1.208(3)    &1.012(6)    &0.994(7)    &0.994(9)    &1.006(10)     & [3.75--4.05]     &0.013\\ 
\tableline
\enddata
\tablecomments{The one sigma limits on $f_\mathrm{esc,rel}$ values are reported as a function of redshift and magnitude bins. 
The average IGM transmission $<T>$ in the middle of the redshift range and 
convolved with the IB VIMOS filter, the intrinsic luminosity ratio $L1500/L_{LyC}$=7 are adopted.
The total $f_\mathrm{esc}$ ($f_\mathrm{esc} = f_\mathrm{esc,rel} \times 10^{-0.4\times A1500}$) can be obtained by assuming 
the average dust absorption of the sample, A1500$\simeq$0.65 (see text).
Within parenthesis the number of sources used in the calculation having magnitude less than the corresponding
column head and belonging to the redshift interval ($redshift$ column). These limits are further decreased by
a factor 1.5 if the recent results of Prochaska et al. (2010) and Songaila \& Cowie (2010) on the LLSs statistics 
are considered (see text).}
\end{deluxetable}

\clearpage 
\begin{deluxetable}{lccccc}
\tabletypesize{\scriptsize}
\tablecaption{Galaxies in the HUDF with photometric redshift in the range 3.4--4.0 and 
27$<$\wi$<$28.5 with a detection in the IB image with S/N$>$2 in the 1.2\see~aperture.
  \label{sample_BPZ}}
\tablewidth{0pt}
\tablehead{
\colhead{ID} & \colhead{GOODS ID} & \colhead{zphot} & \colhead{\wi} &
\colhead{S/N 1.2\see} & \colhead{S/N 2.1\see}
}
\startdata
1 & J033229.90-274721.5 & 3.559 & 28.10 & 4.6 & 7.0\\
2 & J033230.79-274740.6 & 3.495 & 27.63 & 3.6 & 10.1\\
3 & J033232.09-274726.9 & 3.777 & 27.21 & 2.7 & 5.5\\
4 & J033232.83-274630.0 & 3.619 & 27.78 & 3.6 & 11.0\\
5 & J033234.63-274819.4 & 3.487 & 27.78 & 2.4 & 1.8\\
6 & J033236.50-274550.8 & 3.650 & 27.80 & 2.9 & 3.5\\
7 & J033236.67-274802.9 & 3.764 & 27.83 & 2.2 & 7.6\\
8 & J033236.67-274743.4 & 3.681 & 27.85 & 4.7 & 14.0\\
9 & J033236.94-274757.5 & 3.507 & 28.43 & 2.6 & 5.9\\
10 & J033237.87-274552.9 & 3.562 & 27.21 & 3.7 & 7.2\\
11 & J033238.30-274728.7 & 3.488 & 28.15 & 3.8 & 5.6\\
12 & J033238.50-274902.6 & 3.592 & 28.03 & 9.2 & 9.5\\
13 & J033239.43-274956.6 & 3.546 & 28.36 & 2.6 & 5.0\\
14 & J033240.70-274936.8 & 3.830 & 27.05 & 3.1 & 9.4\\
15 & J033240.85-274912.0 & 3.699 & 27.91 & 4.4 & 6.1\\
16 & J033241.33-274548.2 & 3.425 & 28.23 & 4.6 & 9.4\\
17 & J033241.57-274604.1 & 3.497 & 27.93 & 3.8 & 4.2\\
18 & J033241.57-274821.2 & 3.640 & 28.35 & 3.2 & 8.4\\
19 & J033241.83-274811.9 & 3.401 & 27.59 & 2.7 & 5.2\\
20 & J033241.86-274718.2 & 3.509 & 28.38 & 2.9 & 6.6\\
21 & J033242.24-274859.4 & 3.460 & 27.57 & 21.0 & 33.9\\
22 & J033242.77-274618.1 & 3.416 & 28.45 & 2.9 & 5.5\\
23 & J033242.89-274845.7 & 3.761 & 28.19 & 3.5 & 3.2\\
24 & J033244.14-274737.7 & 3.780 & 28.45 & 2.0 & 3.9\\
25 & J033246.03-274752.8 & 3.854 & 27.98 & 2.4 & 2.9\\
26 & J033246.97-274730.5 & 3.610 & 27.15 & 4.4 & 7.3\\
\tableline
\enddata
\end{deluxetable}

\clearpage
\clearpage

\clearpage

\clearpage
\clearpage
\clearpage
\clearpage
\clearpage
\clearpage
\clearpage
\clearpage
\clearpage
\clearpage
\clearpage
\clearpage

\clearpage

\clearpage
\clearpage

\clearpage


\clearpage
\clearpage

\clearpage

\clearpage

\end{document}